\documentclass[sigconf,natbib,anonymous=false,review=false,screen=true]{acmart}

\usepackage{booktabs} 
\usepackage{algorithm, algpseudocode}
\usepackage{amsmath}
\usepackage{graphics}
\usepackage{epsfig}
\usepackage{graphicx}
\usepackage{xcolor}
\usepackage[skip=0pt]{caption}
\usepackage{subfigure}
\usepackage{balance}
\usepackage{multirow}
\usepackage{mathrsfs}
\usepackage{acronym}
\usepackage{placeins}
\usepackage{xcolor}
\usepackage[inline]{enumitem}

\AtBeginDocument{%
  \providecommand\BibTeX{{%
    \normalfont B\kern-0.5em{\scshape i\kern-0.25em b}\kern-0.8em\TeX}}}

\theoremstyle{definition}
\newtheorem{definition}{Definition}[section]

\acrodef{KGE}{knowledge graph embedding}
\acrodef{KGC}{knowledge graph completion}
\acrodef{OOKB}{out-of-knowledge-base}
\acrodef{OOKG}{out-of-knowledge-graph}
\acrodef{KG}{knowledge graph}
\acrodef{KGs}{knowledge graphs}
\acrodef{GNN}{graph neural network}
\acrodef{VN}{virtual neighbor}
\acrodef{VNC}{virtual neighbor network with inter-rule correlations}

\newcommand{\header}[1]{\vspace*{1mm}\noindent\textbf{#1.}}

\copyrightyear{2023}
\acmYear{2023}
\setcopyright{acmlicensed}
\acmConference[CIKM '23] {Proceedings of the 32nd ACM International Conference on Information and Knowledge Management}{October 21--25, 2023}{Birmingham, United Kingdom.}
\acmBooktitle{Proceedings of the 32nd ACM International Conference on Information and Knowledge Management (CIKM '23), October 21--25, 2023, Birmingham, United Kingdom}
\acmPrice{15.00}
\acmISBN{979-8-4007-0124-5/23/10}
\acmDOI{10.1145/3583780.3614938}

\author{Zihan Wang}
\orcid{0000-0003-0493-2668}
\affiliation{
\institution{Shandong University}
\institution{University of Amsterdam}
\city{}
\country{}
}
\email{zihanwang.sdu@gmail.com}

\author{Kai Zhao}
\orcid{0000-0003-1040-0211}
\affiliation{
\institution{Georgia State University}
\city{}
\country{}
}
\email{kzhao4@gsu.edu}

\author{Yongquan He}
\orcid{0000-0002-3079-8530}
\affiliation{%
  \institution{Meituan}
  \city{}
  \country{}
}
\email{heyongquan@meituan.com}

\author{Zhumin Chen}
\orcid{0000-0003-4592-4074}
\affiliation{\institution{Shandong University}
\city{}
\country{}
}
\email{chenzhumin@sdu.edu.cn}

\author{Pengjie Ren}
\orcid{0000-0003-2964-6422}
\affiliation{\institution{Shandong University}
\city{}
\country{}
}
\email{jay.ren@outlook.com}

\author{Maarten	de Rijke}
\orcid{0000-0002-1086-0202}
\affiliation{\institution{University of Amsterdam}
\city{}
\country{}
}
\email{m.derijke@uva.nl}

\author{Zhaochun Ren}
\orcid{0000-0002-9076-6565}
\authornote{Corresponding author}
\affiliation{
\institution{Leiden University}
\city{}
\country{}
}
\email{zhc.ren@gmail.com}

\begin{document}

\title{Iteratively Learning Representations for Unseen Entities with Inter-Rule Correlations}


\begin{abstract}
Recent work on \acf{KGC} focuses on acquiring embeddings of entities and relations in knowledge graphs. 
These embedding methods necessitate that all test entities be present during the training phase, resulting in a time-consuming retraining process for \acf{OOKG} entities.
To tackle this predicament, current inductive methods employ \acp{GNN} to represent unseen entities by aggregating information of the known neighbors, and enhance the performance with additional information, such as attention mechanisms or logic rules. 
Nonetheless, Two key challenges continue to persist:
\begin{enumerate*}[label=(\roman*)]
    \item identifying inter-rule correlations to further facilitate the inference process, and
    \item capturing interactions among rule mining, rule inference, and embedding to enhance both rule and embedding learning.
\end{enumerate*}

In this paper, we propose a \acf{VNC} to address the above challenges.
\ac{VNC} consists of three main components:
\begin{enumerate*}[label=(\roman*)]
\item rule mining, 
\item rule inference, and 
\item embedding. 
\end{enumerate*}
To identify useful complex patterns in knowledge graphs, both logic rules and inter-rule correlations are extracted from knowledge graphs based on operations over relation embeddings.
To reduce data sparsity, virtual networks for \ac{OOKG} entities are predicted and assigned soft labels by optimizing a rule-constrained problem.
We also devise an iterative framework to capture the underlying interactions between rule and embedding learning.
Experimental results on both link prediction and triple classification tasks show that the proposed \ac{VNC} framework achieves state-of-the-art performance on four widely-used knowledge graphs. 
Our code and data are available at \url{https://github.com/WZH-NLP/OOKG}.
\end{abstract}



\maketitle

\acresetall


\section{Introduction}
\Acp{KG} are widely used to store structured information and facilitate a broad range of downstream applications, such as question answering~\citep{DBLP:conf/acl/LanJ20,DBLP:journals/tois/ZhangWMH19}, dialogue systems~\citep{DBLP:conf/acl/HeBEL17}, recommender systems~\citep{DBLP:conf/kdd/Wang00LC19,DBLP:journals/tois/WangZWZLXG19, mu2021knowledge}, and information extraction~\citep{DBLP:conf/emnlp/WangZWZCZZC18,DBLP:journals/tois/YangHSJLN21}. 
A typical \ac{KG} represents facts as triples in the form of (\emph{head entity}, \emph{relation}, \emph{tail entity}), e.g., (\emph{Alice}, \emph{IsBornIn}, \emph{France}). 
Despite their size, \acp{KG} suffer from  incompleteness~\citep{DBLP:conf/naacl/MinGW0G13}. 
Therefore, \acf{KGC}, which is aimed at automatically predicting missing information, is a fundamental task for \acp{KG}.
To address the \ac{KGC} task, \acf{KGE} methods have been proposed and attracted increasing attention~\citep{DBLP:conf/aaai/BordesWCB11,DBLP:conf/nips/BordesUGWY13, DBLP:conf/esws/SchlichtkrullKB18, DBLP:conf/aaai/GuoWWWG18,DBLP:conf/acl/WangWGD18, DBLP:conf/aaai/NiuZ0CLLZ20,DBLP:conf/www/ZhangPWCZZBC19}. 

Previous \ac{KGE} methods focus on transductive settings, requiring all entities to be observed during training. 
In real-world scenarios, however, \acp{KG} evolve dynamically since \acf{OOKG} entities emerge frequently~\citep{graus-birth-2018}. 
For example, about 200 new entities are added to DBPedia on a daily basis~\cite{DBLP:conf/aaai/ShiW18}. 
Fig.~\ref{Fig: OOKG problem} shows an example of the \ac{OOKG} entity problem. 
Given the observed \ac{KG}, ``sun'' is the newly added entity and there exists the auxiliary connection between ``sun'' and the known entity (i.e., (\emph{sun}, \emph{surroundedBy}, \emph{planets})). Based on observed and auxiliary facts, our goal is to embed \ac{OOKG} entities and predict missing facts (e.g., (\emph{sun}, \emph{attract}, \emph{mass})). 
So far, to represent newly emerging entities, a time-consuming retraining process over the whole \ac{KG} is unavoidable for most conventional embedding methods. 
To address this issue, an inductive \ac{KGE} framework is needed.

Some previous work~\citep{DBLP:conf/ijcai/HamaguchiOSM17,DBLP:conf/aaai/WangHLP19,DBLP:journals/corr/abs-2009-12765} represents \ac{OOKG} entities using their observed neighborhood structures.
These frameworks suffer from a data sparsity problem~\citep{DBLP:conf/cikm/HeWZTR20,DBLP:conf/cikm/ZhangW0XLZ21}.
To address this sparsity issue, GEN~\citep{DBLP:conf/nips/BaekLH20} and HRFN~\citep{DBLP:conf/cikm/ZhangW0XLZ21} combine meta-learning frameworks with \acp{GNN} to simulate unseen entities during meta-training.
But they utilize triples between unseen entities, which may be missing or extremely sparse in real-world scenarios.
The VN network~\citep{DBLP:conf/cikm/HeWZTR20} alleviates the sparsity problem by inferring additional \acfp{VN} of the \ac{OOKG} entities with logic rules and symmetric path rules.

Despite these advances, current inductive knowledge embedding methods face the following two challenges:

\begin{figure*}
  \centering
    \includegraphics[width=0.9\linewidth]{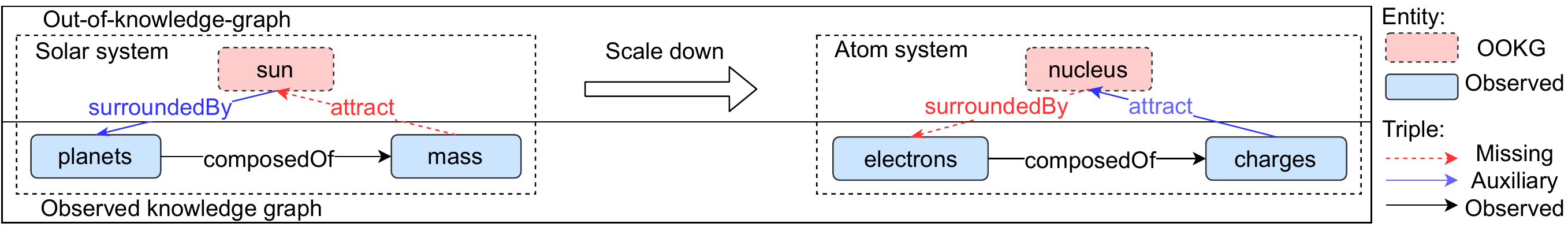}
    \caption{An example of the \ac{OOKG} entity problem. 
    Our aim is to predict missing facts of the \ac{OOKG} entities.
    }
    \label{Fig: OOKG problem}
\end{figure*}

\header{Challenge 1: Identifying inter-rule correlations}
Previous methods for inductive knowledge embedding mainly focus on modeling one or two hop local neighborhood structures, or mining rules for the \ac{OOKG} entities.
Other complex patterns helpful for the predictions of missing facts, such as inter-rule correlations, are ignored. 
As shown in Fig.~\ref{Fig: OOKG problem}, the extracted logic rule $(\mathit{sun}, \mathit{surroundedBy}, \mathit{planets})$ ${}\land(\mathit{planets}, \mathit{composedOf}, \mathit{mass})\to(\mathit{sun}$, $\mathit{attract}, \mathit{mass})$ describes the principle of the solar system. 
Given the fact that the solar system, and atom system are correlated (since the ``\emph{nucleus}'' is the scale-down ``\emph{sun}'' in the atom), the missing fact $(\mathit{nucleus}, \mathit{Surround}$- $\mathit{edBy}, \mathit{electrons})$ and new rule $(\mathit{nucleus}, \mathit{surroundedBy}, \mathit{electrons})\land(\mathit{electrons}, \mathit{composedOf}, \mathit{charges})\to(\mathit{nucleus}, \mathit{attract}, \mathit{charges})$ are obtained easily through the analogy between the solar and atom system.
In this work, such correlations are extracted and modeled to facilitate inductive \ac{KGE} methods.
By identifying inter-rule correlations, our proposed method is able to discover most (more than 80\%) of symmetric path (SP) rules used by VN network~\citep{DBLP:conf/cikm/HeWZTR20} and other useful patterns in \ac{KGs} to further improve embedding learning (see \S\ref{subsubsec:Relations to symmetric path rules}). 

\header{Challenge 2: Capturing the interactions among rule mining, rule inference, and embedding}
\sloppy LAN~\citep{DBLP:conf/aaai/WangHLP19} utilizes constant logic rule confidences to measure neighboring relations' usefulness, while VN network~\citep{DBLP:conf/cikm/HeWZTR20} employs the heuristic rule mining method (i.e., AMIE+~\citep{DBLP:journals/vldb/GalarragaTHS15}). 
In that case, prior work fails to capture interactions among rule mining, rule inference, and embedding.
In fact, these three processes (i.e., rule mining, rule inference, and embedding) benefit and complement each other.
Specifically, rules can infer missing facts more accurately with refined embeddings, while predicted facts help to learn the embeddings of higher quality~\citep{DBLP:conf/aaai/GuoWWWG18}. 
Besides, rule learning using \ac{KG} embeddings can transform the mining process from discrete graph search into calculations in continuous spaces, reducing the search space remarkably~\citep{DBLP:conf/www/ZhangPWCZZBC19}. 
In this work, we design an iterative framework for rule mining, rule inference, and embedding to incorporate the relations among the above three stages, as Fig.~\ref{fig: VNC overview} illustrates.

\vspace*{1mm}\noindent%
To address the two challenges listed above, we propose an inductive knowledge embedding framework, named \acfi{VNC}, to iteratively infer virtual neighbors for the \ac{OOKG} entities with logic rules and inter-rule correlations. As Fig.~\ref{fig: VNC overview} illustrates, VNC is composed of three main stages: 
\begin{enumerate*}[label=(\roman*)]
\item rule mining, 
\item rule inference, and 
\item embedding. 
\end{enumerate*}
In the rule mining process, to capture useful complex patterns in \ac{KG}, both logic rules and inter-rule correlations are extracted from \acp{KG}, and assigned confidence scores via calculations over relation embeddings. 
To alleviate the data sparsity problem, \acfp{VN} of entities are inferred utilizing the deductive capability of rules. 
By solving a convex rule-constrained problem, soft labels of \acp{VN} are optimized. 
Next, the \ac{KG} with softly predicted \acp{VN} is input to the \ac{GNN}-based encoder, which consists of structure-aware and query-aware layers.
Moreover, entity embeddings obtained by aggregating neighbors in the encoder are taken as the initialization for the embedding-based decoder. Finally, optimal entity and relation embeddings are derived by minimizing the global loss over observed and softly labeled fact triples.
The above three processes are conducted iteratively during training.

Our contributions can be summarized as follows:
\begin{enumerate*}[label=(\roman*)]
    \item We propose an inductive knowledge embedding paradigm, named \ac{VNC}, to address the \ac{OOKG} entity problem.
    \item We develop an embedding-enhanced rule mining scheme to identify logic rules and inter-rule correlations simultaneously. 
    \item We design an iterative framework to explore the interactions among rule mining, rule inference, and embedding.
    \item Experimental results show that the proposed \ac{VNC} achieves state-of-the-art performance in both link prediction and triple classification tasks. 
\end{enumerate*}

\section{Related Work}
\label{sec:Related works}

\textbf{Knowledge graph completion.}
Knowledge graph completion (KGC) methods have been extensively studied and mainly fall under the embedding-based paradigm~\citep{DBLP:conf/icml/TeruDH20,DBLP:journals/tacl/WangGZZLLT21}. 
The aim of \acf{KGE} methods is to map the entities and relations into continuous vector spaces and then measure the plausibility of fact triples using score functions. 
Early work designs shallow models solely relying on triples in the observed \ac{KG}s~\citep{DBLP:conf/aaai/BordesWCB11,DBLP:conf/nips/BordesUGWY13,DBLP:conf/icml/NickelTK11}. 
One line of recent works focus on devising more sophisticated triple scoring functions, including TransH~\citep{DBLP:conf/aaai/WangZFC14}, TransR~\citep{DBLP:conf/aaai/LinLSLZ15}, RotatE~\citep{DBLP:conf/iclr/SunDNT19}, DistMult~\citep{DBLP:journals/corr/YangYHGD14a}, and  Analogy~\citep{DBLP:conf/icml/LiuWY17}.
Another line of recent methods is to incorporate useful information beyond triples, including relation paths~\citep{DBLP:conf/acl/NeelakantanRM15,DBLP:conf/ecir/ZhangWXLS18} and logic rules~\citep{DBLP:conf/aaai/GuoWWWG18,DBLP:conf/acl/WangWGD18, DBLP:conf/aaai/NiuZ0CLLZ20,DBLP:conf/www/ZhangPWCZZBC19}.
Besides, deep neural network based methods~\citep{DBLP:conf/aaai/DettmersMS018,DBLP:conf/ijcai/WangRHZH19, DBLP:conf/www/WangWSWNAXYC21} and language model based methods~\citep{DBLP:journals/tacl/WangGZZLLT21,DBLP:conf/acl/0046ZWL22} also show promising performance.

\header{Inductive knowledge embedding}
Despite the success in \ac{KGC} problem, the above \ac{KGE} methods still focus on the transductive settings, requiring all the test entities to be seen during training.
Motivated by the limitations of traditional \ac{KGE} methods,
recent works~\citep{DBLP:conf/ijcai/HamaguchiOSM17,DBLP:conf/aaai/WangHLP19,DBLP:journals/corr/abs-2009-12765,DBLP:conf/nips/BaekLH20} take the known neighbors of the emerging entities as the inputs of inductive models. \citet{DBLP:conf/ijcai/HamaguchiOSM17} employ the \acf{GNN} and aggregate the pretrained representations of the existing neighbors for unseen entities. 
To exploit information of redundancy and query relations in the neighborhood, LAN~\citep{DBLP:conf/aaai/WangHLP19} utilizes a logic attention network as the aggregator. 
GEN~\citep{DBLP:conf/nips/BaekLH20} and HRFN~\citep{DBLP:conf/cikm/ZhangW0XLZ21} design meta-learning frameworks for \ac{GNN}s to simulate the unseen entities during meta-training.
However, they utilize unseen-to-unseen triples, which are unavailable in the \ac{OOKG} entity problem.
VN network~\citep{DBLP:conf/cikm/HeWZTR20} alleviates the data sparsity problem by inferring virtual neighbors for the \ac{OOKG} entities.
In addition, InvTransE and InvRotatE~\citep{DBLP:journals/corr/abs-2009-12765} represent \ac{OOKG} entities with the optimal estimations of translational assumptions.
Another type of inductive methods represent unseen entities via learning entity-independent semantics, including rule based~\citep{DBLP:conf/nips/SadeghianADW19} and GNN based~\citep{DBLP:conf/icml/TeruDH20,DBLP:conf/aaai/ChenHWW21} methods. 
However, the above methods focus on a different inductive \ac{KGC} task (i.e., completing an entirely new \ac{KG} during testing), and are not able to take advantage of embeddings of known entities or inter-rule correlations. 
In our experiments, we also conduct a comprehensive comparison between our proposed model and entity-independent methods (see \S\ref{sec:Comparisions with entity-independent methods}).

The most closely related work is MEAN~\citep{DBLP:conf/ijcai/HamaguchiOSM17}, LAN~\citep{DBLP:conf/aaai/WangHLP19}, VN network~\citep{DBLP:conf/cikm/HeWZTR20}, InvTransE and InvRotatE~\citep{DBLP:journals/corr/abs-2009-12765}. 
These previous inductive embedding methods ignore inter-rule correlations, and do not capture interactions among rule mining, rule inference, and embedding. 
In our proposed model \ac{VNC}, 
to model useful complex patterns in graphs, logic rules and inter-rule correlations are identified simultaneously. 
We design an iterative framework to incorporate interactions among rule mining, rule inference, and embedding.


\section{Definitions}
\label{sec:Definition}

\begin{definition}[Knowledge graph]
A \emph{knowledge graph} $\mathcal{K}$ can be regarded as a multi-relational graph, consisting of a set of observed fact triples, i.e.,
$\mathcal{O} = \{t_{o}\}$,  where  $t_{o} = (e_{i}, r_{k}, e_{j})$. 
Each fact triple consists of two entities $e_{i}, e_{j} \in \mathcal{E}_{o}$, 
and one type of relation $r_{k} \in \mathcal{R}$, where $\mathcal{E}_{o}$ and $\mathcal{R}$ are the entity and relation sets respectively. 
For each triple $(e_{i}, r_{k}, e_{j}) \in \mathcal{K}$, we denote the reverse version of relation $r_{k}$ as $r^{-1}_{k}$ and add $(e_{j}, r^{-1}_{k}, e_{i})$
to the original \ac{KG} $\mathcal{K}$.
\end{definition}

\begin{definition}[Out-of-knowledge-graph entity problem] Following~\citep{DBLP:conf/ijcai/HamaguchiOSM17, DBLP:journals/corr/abs-2009-12765}, we formulate the \acfi{OOKG} entity problem as follows.
The auxiliary triple set $\mathcal{AUX}$ contains the unseen entities $\mathcal{E}_{u} = \mathcal{E}_{aux}/\mathcal{E}_{o}$,
and each triple in $\mathcal{AUX}$ contains exactly one \ac{OOKG} entity and one observed entity.
And $\mathcal{O}$ is observed during training,
while the auxiliary triple set $\mathcal{AUX}$ 
connecting \ac{OOKG} and observed entities is only accessible at test time.  
Note that, no additional relations are involved in $\mathcal{AUX}$.
Given $\mathcal{AUX}$ and $\mathcal{O}$, the goal is to correctly identify missing fact triples that involve the \ac{OOKG} entities.
\end{definition}

\begin{definition}[Logic rules]
For logic rules, following~\citep{DBLP:conf/aaai/GuoWWWG18, DBLP:conf/acl/WangWGD18}, we consider a set of first-order logic rules with different confidence values for a give \ac{KG}, represented as $\mathcal{F}^\mathit{logic}=\{(f^\mathit{logic}_{m}, \lambda^\mathit{logic}_{m})\}_{m=1}^M$, where $f^\mathit{logic}_{m}$ is the $m$-th logic rule. $\lambda^\mathit{logic}_{m} \in [0, 1]$ denotes its confidence value, and rules with higher confidence values are more likely to hold. 
Here, $f^\mathit{logic}_{m}$ is in the form of $\mathit{body} \rightarrow \mathit{head}$. 
In this paper, we restrict rules to be Horn clause rules, where the rule head is a single atom, and the rule body is a conjunction of one or more atoms.
For example, such kind of logic rule can be:
\begin{equation}
\label{: an example of logic rule}
    (x, \mathit{surroundedBy}, y)\land (y, \mathit{composedOf}, z) \rightarrow (x, \mathit{attract}, z),
\end{equation}
where $x$, $y$, $z$ are entity variables. 
Similar to previous rule learning work~\citep{DBLP:conf/aaai/GuoWWWG18, DBLP:conf/www/GalarragaTHS13}, we focus on closed-path (CP) rules to balance the expressive power of mined rules and the efficiency of rule mining. 
In a CP rule, the sequence of triples in the rule body forms a path from the head entity variable to the tail entity variable of the rule head.
By replacing all variables with concrete entities in the given \ac{KG}, we obtain a grounding of the rule.
For logic rule $f^\mathit{logic}_{m}$, we denote the set of its groundings as $\mathcal{G}_{m}^\mathit{logic} = \{g^\mathit{logic}_{mn}\}_{n=1}^{N_{m}}$.
\end{definition}

\begin{definition}[Inter-rule correlations]
In addition to logic rules, we also consider a set of inter-rule correlations with different confidence levels, denoted as $\mathcal{F}^\mathit{corr}=\{(f^\mathit{corr}_{v}, \lambda^\mathit{corr}_{v})\}_{v=1}^{V}$, where $f^\mathit{corr}_{v}$ is the $v$-th inter-rule correlation and $\lambda^\mathit{corr}_{v}$ is the corresponding confidence value.
Based on the logic rule $f^\mathit{logic}_{m}$, we define the corresponding inter-rule correlations as:
\begin{equation}
f^{\mathit{corr}}_{v_{\mathit{mpq}}}: f^{\mathit{logic}}_{m}\xrightarrow{\mathit{path}_{q}\left(f^{\mathit{logic}}_{m}, f'^{\mathit{logic}}_{\mathit{mp}}\right)} f'^{\mathit{logic}}_{\mathit{mp}},
\label{Eq: inter-rule correlation}
\end{equation}
where $f'^{logic}_{mp}$ is the $p$-th ``incomplete'' logic rule in the same form as  $f^{logic}_{m}$ but with one missing triple in the rule body.
For example, as Fig.~\ref{Fig: OOKG problem} shows, the rule for the atom system are incomplete since $(x, \mathit{surroundedBy}, y)$ in the logic rule $(x, \mathit{surroundedBy}, y)\land (y, \mathit{composedOf}, z) \rightarrow (x, \mathit{attract}, z)$ is missing. 
Note that the rules with only rule head missing are not regarded as the ``incomplete'' rules, because the missing rule head can be directly inferred by extracted logic rules. 
The $q$-th inter-rule path between the logic rule  $f^\mathit{logic}_{m}$ and incomplete rule $f'^{\mathit{logic}}_\mathit{mp}$ is represented as follows:
$
\small
\mathit{path}_{q}(f^\mathit{logic}_{m}, f'{^\mathit{logic}}_\mathit{mp}) : (x_{1}, r_{1}, x_{2})\land{}  (x_{2}, r_{2}, x_{3})\land {} 
\cdots\land(x_{k},r_{k}, x_{k+1}),
$
where $r_{i} \in \mathcal{R}$ denotes a relation in \ac{KG} and $x_{i}$ is the entity variable. 
To represent the correlations between rules, we assume that the inter-rule path only exists between entities of the same position in two rules. 
For example, in Fig.~\ref{Fig: OOKG problem},
the inter-rule path is $(\mathit{sun}, \mathit{scaleDown}, \mathit{nucleus})$ indicating that the nucleus is the scaled-down sun in the atom system.
Similar to the logic rules, we obtain the set of groundings $\mathcal{G}^\mathit{corr}_{v} = \{g^\mathit{corr}_\mathit{vw}\}_{w=1}^{W_{v}}$ for $f^\mathit{corr}_{v}$ by replacing variables with concrete entities.

\end{definition}

\begin{definition}[Virtual neighbors]
To address the data sparsity problem, we introduce virtual neighbors into the original \ac{KG}. 
As mentioned above, virtual neighbors are inferred by the extracted rules (i.e., logic rules and inter-rule correlations). 
Specifically, if a triple $(e'_{i}, r'_{k}, e'_{j})$ inferred by rules does not exist in either the observed triple set $\mathcal{O}$ or auxiliary triple set $\mathcal{AUX}$, we suppose that $e'_{i}$ and $e'_{j}$ are the virtual neighbors to each other.
In our paper, we denote the set containing such kind of triples as $\mathcal{VN} = \{t_{vn}\}$, where $t_{vn}$ is a triple with the virtual neighbors. 
\end{definition}


\section{Method}
\label{sec:Method}
\begin{figure*}[t] 
    \centering
    \includegraphics[width=0.85\linewidth]{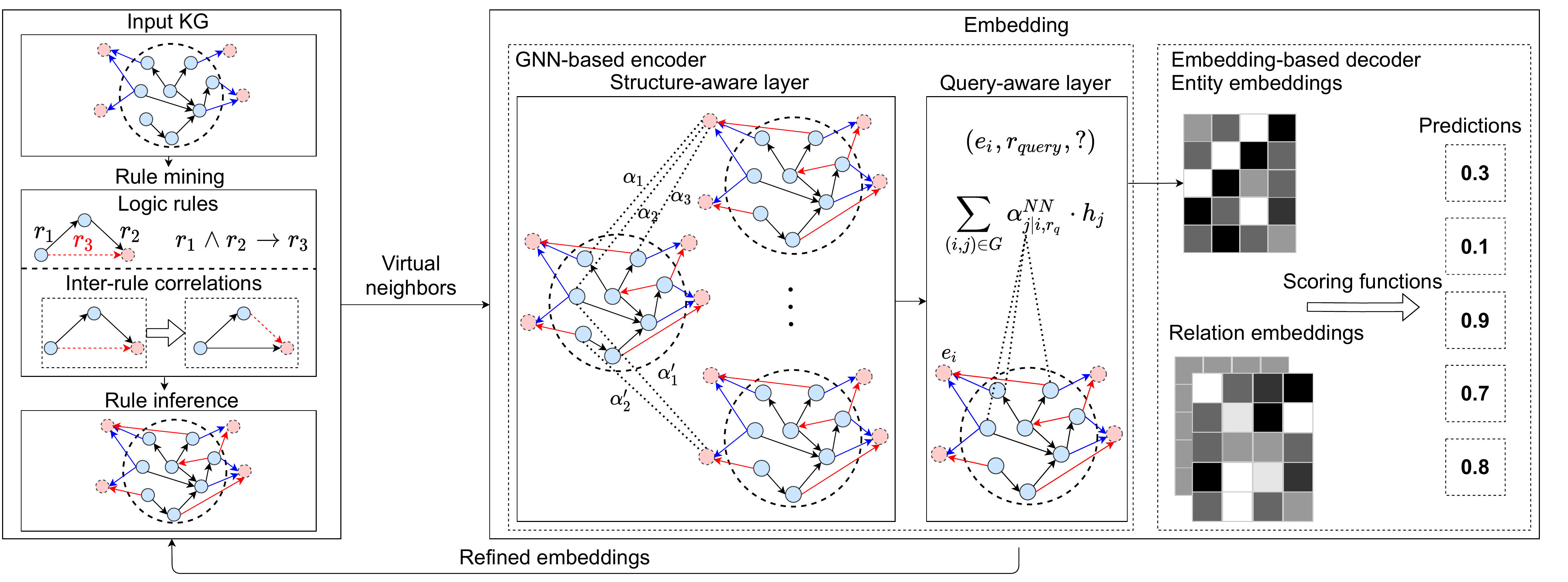}
    \caption{An overview of \ac{VNC}. 
    VNC has three main stages: rule mining, rule inference, and embedding.
    }
    \label{fig: VNC overview}
\end{figure*}

In this section, we describe the \ac{VNC} framework, 
our proposed method for the \ac{OOKG} entity problem.
As illustrated in Fig.~\ref{fig: VNC overview}, the framework has three stages: rule mining (\S\ref{subsec: rule mining}), rule inference (\S\ref{subsec: rule inference}), and embedding (\S\ref{subsec: embedding}). 
In the rule mining stage, Given the knowledge graph, the rule pool is first generated by searching plausible paths, and confidence values are calculated using the current relation embeddings $\mathbf{R}$. 
Then, in the rule inference stage, a new triple set with virtual neighbors $\mathcal{VN} = \{t_{\mathit{vn}}\}$ is inferred from rule groundings. And each predicted triple $t_{\mathit{vn}}$ is assgined a soft label $s(t_{\mathit{vn}})\in[0, 1]$ by solving a rule-constrained optimization problem.
The knowledge graph with virtual neighbors is inputted into GNN-based encoder consisting of both structure and query aware layers. 
Next, with the entity embeddings $\mathbf{E} = \mathbf{H}^{O}$, where $\mathbf{H}^{O}$ is the output of \ac{GNN} layers, the embedding-based decoder projects relations into embeddings $\mathbf{R}$ and calculate the truth level $\phi(\cdot)$ for each fact triple as follows (take DistMult~\citep{DBLP:journals/corr/YangYHGD14a} as an example):
\begin{equation}
\phi(e_{i}, r_{k}, e_{j}) = \mathbf{e}_{i}^{T}\mathbf{R}_{k}\mathbf{e}_{j},
\label{eq:DistMult}
\end{equation}
where $\mathbf{e}_{i}, \mathbf{e}_{j}$ are the normalized entity embeddings for entity $e_{i}$ and $e_{j}$ respectively, and $\mathbf{R}_{k}$ is a diagonal matrix for relation $r_{k}$.
These three stages are conducted iteratively during training (see \S\ref{subsec: training algorithm}).


\subsection{Rule mining}
\label{subsec: rule mining}
Given the observed knowledge graph, 
rule mining stage first generates a pool of logic rules by finding possible paths. 
Then, based on the complete logic rules, inter-rule correlations are discovered by searching incomplete rules and inter-rule paths.
Finally, the confidence values are computed using relation embeddings. 

\subsubsection{Rule pool generation}
Before computing confidence scores, rules should be extracted from the observed \ac{KG}.

For logic rules, we are only interested in closed-path (CP) rules.
Therefore, given the rule head, the search for candidate logic rules is reduced to finding plausible paths for rule bodies.
Specifically, one of fact triples in the observed \ac{KG} $\mathcal{K}$ (e.g., $(e_{1}, r, e_{2}) \in \mathcal{O}$) is first taken as the candidate rule head, and then the possible paths between the head entity and tail entity of the rule head (e.g., $(e_{1}, r_{1}, e_{3})\land (e_{3}, r_{2}, e_{2})$) is extracted.
In this way, the candidate logic rule $(x, r_{1}, z)\land (z, r_{2}, y)\rightarrow (x, r, y)$ is induced from the given \ac{KG}.
For computational efficiency, we restrict the length of paths in rule bodies to at most 2 (i.e., the length of rules is restricted to at most 3).
Note that, there may still exist numerous redundant and low quality rules in the above extraction process. Therefore, following~\citep{DBLP:conf/www/GalarragaTHS13, DBLP:conf/www/ZhangPWCZZBC19}, further filtering is conducted, and only rules with $\mathit{support} > 1$, $\mathit{head}\ \mathit{coverage} > \alpha_{HC}$, and $\mathit{standard}\ \mathit{confidence} > \alpha_{SC}$
are selected, where $\alpha_{HC}$ and $\alpha_{SC}$ are preset thresholds.

Based on mined logic rules, there are two steps for generating possible inter-rule correlations:
\begin{enumerate*}[label=(\roman*)]
\item \textbf{Finding incomplete rules}. To this end, our aim is to identify all the ``incomplete'' rules for the mined logic rules. 
Specifically, given the $m$-th logic rule $f^\mathit{logic}_{m}$ in $\mathcal{K}$, a set of
``incomplete'' logic rules $\{f'^{\mathit{logic}}_{mp}\}$ in the same form as $f^{\mathit{logic}}_{m}$ but with one missing triple in the rule body is recognized in this step.
For example, for the logic rule of length 2 (e.g., $(x, r_{1}, y)\rightarrow (x, r, y)$), there exists only one ``incomplete'' logic rule (e.g., $(x, r_{1}, y)\rightarrow (x, r, y)$ with $(x, r_{1}, y)$ missing). 

\item \textbf{Searching plausible inter-rule paths}. To extract inter-rule paths, we first obtain groundings of logic rules and ``incomplete'' rules by replacing variables with concrete entities. 
For example, a grounding of logic rule $(x, r_{1}, z)\land (z, r_{2}, y)\rightarrow (x, r, y)$ can be $(e_{1}, r_{1}, e_{2})\land (e_{2}, r_{2}, e_{3})$ $\rightarrow (e_{1}, r, e_{3})$,  and a grounding of the corresponding ``incomplete'' rules can be $(e'_{1}, r_{1}, e'_{2})\land (e'_{2}, r_{2}, e'_{3})\rightarrow (e'_{1}, r, e'_{3})$, where $e_{i},\ e'_{i}\in \mathcal{E}_{o}$ and $(e'_{1}, r_{1}, e'_{2}) \notin \mathcal{O}$.
Then, the paths between entities of the same position in logic and ``incomplete'' rules (e.g., paths between $e_{1}$ and $e'_{1}$) are searched. 
Here, we estimate the reliability of inter-rule paths using the path-constraint resource allocation (PCRA) algorithm~\citep{DBLP:conf/emnlp/LinLLSRL15}, and keep paths with $\mathit{Reliability} > \alpha_{PCRA}$, where $\alpha_{PCRA}$ is the threshold for the path reliability.
For computational efficiency, the length of inter-rule paths is limited to at most 3.
\end{enumerate*}
In the next step, similar to mining logic rules, we filter out inter-rule correlations of low quality with $\mathit{support}$, $\mathit{head}\ \mathit{coverage}$, and $\mathit{standard}\ \mathit{confidence}$.

\subsubsection{Confidence computation}
Given the generated rule pool and current relation embedding $\mathbf{R}$, the confidence computation assigns a score $\lambda_{m}$ for each extracted rule $f_{m}$.

For each logic rule, the rule body and rule head can be considered as two associated paths.
Inspiring by previous works~\citep{DBLP:conf/www/ZhangPWCZZBC19, DBLP:conf/ijcai/OmranWW18}, the confidence level of each logic rule can be measured by the similarity between paths of the rule body and rule head.
To be specific, suppose the path of the rule body $\mathit{path}_\mathit{body}: (x_{1}, r_{1}, x_{2})\land (x_{2}, r_{2}, x_{3})\land \cdots\land(x_{k},r_{k}, x_{k+1})$, and the path of the rule head $\mathit{path}_\mathit{head}: (x_{1}, r, x_{k+1})$, the corresponding confidence level $\lambda^{logic}_{m}$
is calculated as follows:
\begin{equation}
\lambda^\mathit{logic}_{m} ={\rm sim}(\mathbf{path}_\mathit{body}, \mathbf{path}_\mathit{head}),
\label{Eq: similarity funciton}
\end{equation}
where $\mathbf{path}_\mathit{body}$ and $\mathbf{path}_\mathit{head}$ are embeddings for the paths of the rule body and rule head, respectively. And ${\rm sim}(\cdot)$ is the similarity function. 
In \ac{VNC}, we consider a variety of methods based on translating or bilinear operations in the embedding stage. 
Thus, we define two kinds of similarity functions and path representations for different embedding methods.
For translational decoders (e.g., TransE), the path representations and similarity function in Eq.~\ref{Eq: similarity funciton} are defined as follows:
\begin{equation}
\label{Eq: transition-based similarity}
\begin{split}
    \mathbf{path}_{\mathit{body}}& = \mathbf{r}_{1} + \mathbf{r}_{2} + \cdots + \mathbf{r}_{k},\ \mathbf{path}_{\mathit{head}} =  \mathbf{r},\\
    \lambda^{\mathit{logic}}_{m} & =  ||\mathbf{path}_{\mathit{body}} - \mathbf{path}_{\mathit{head}}||_{2},\\
\end{split}
\end{equation}
where $\mathbf{r}_{i}$ and $\mathbf{r}$ are vector embeddings for relation $r_{i}$ and $r$, and similarity function is defined by the $L_{2}$-norm.
For bilinear decoders (e.g., DistMult), the path representations and similarity function in Eq.~\ref{Eq: similarity funciton} are defined as follows:
\begin{equation}
\label{Eq: bilinear similarity}
\begin{split}
    \mathbf{path}_{\mathit{body}}& = \mathbf{M}_{r_{1}} + \mathbf{M}_{r_{2}} + \cdots + \mathbf{M}_{r_{k}},
    \\ 
    \mathbf{path}_{\mathit{head}} & =  \mathbf{M}_{r},
    \\
    \lambda^{logic}_{m} & =  ||\mathbf{path}_{\mathit{body}} - \mathbf{path}_{\mathit{head}}||_{F},\\
\end{split}
\end{equation}
where $\mathbf{M}_{r_{i}}$ and $\mathbf{M}_{r}$ are matrix embeddings for relation $r_{i}$ and $r$, and similarity function is defined by the Frobenius norm. 

On this basis, we calculate confidence scores of inter-rule correlations.
Specifically, for inter-rule correlation in Eq.~\ref{Eq: inter-rule correlation}, we consider the confidences of logic rule and ``incomplete'' rule simultaneously, and define the confidence level $\lambda^\mathit{corr}_{v_\mathit{mpq}}$ for inter-rule correlation $f^\mathit{corr}_{v_\mathit{mpq}}$ as follows:
\begin{equation}
\lambda^\mathit{corr}_{v_\mathit{mpq}} = \lambda^\mathit{logic}_{m} \cdot \lambda'^{\mathit{logic}}_\mathit{mp},
\label{Eq: inter-rule confidence}
\end{equation}
where $\lambda^\mathit{logic}_{m}$ and $\lambda'^{\mathit{logic}}_{mp}$ are the confidences of logic rule $f^\mathit{logic}_{m}$ and ``incomplete'' rule $f'^{\mathit{logic}}_{mp}$, respectively. 
The confidence $\lambda'^{\mathit{logic}}_{mp}$ of ``incomplete'' rule $f'^{\mathit{logic}}_{mp}$ is regarded as the probability of inferring the missing triple using the observed path (from the head entity to the tail entity of the missing triple).
For example, the confidence $\lambda'^{\mathit{logic}}_{mp}$ for ``incomplete'' rule $f'^{\mathit{logic}}_{mp}: (x_{1}, r_{1}, x_{2})\land (x_{2}, r_{2}, x_{3})\rightarrow (x_{1}, r, x_{3})$ (with $(x_{1}, r_{1}, x_{2})$ missing) is computed as (for bilinear decoders):
$
    \lambda'^{\mathit{logic}}_{mp}  =  \|\mathbf{M}_{r} + \mathbf{M}_{r^{-1}_{2}} - \mathbf{M}_{r_{1}}\|_{F},
$
where $\mathbf{M}_{r^{-1}_{i}}$ is the matrix embedding for the reverse version of the relation $r_{i}$. 
Since unreliable paths are filtered out during rule pool generation, the reliability of inter-rule path is not considered.

\subsubsection{Discussion: relation to symmetric path rules}
\label{subsubsec:Relations to symmetric path rules}
\begin{figure}
  \centering
  \subfigure[Symmetric path rule.]{
    \label{Fig: An example of the symmetric path rule.}    \includegraphics[width=0.48\linewidth]{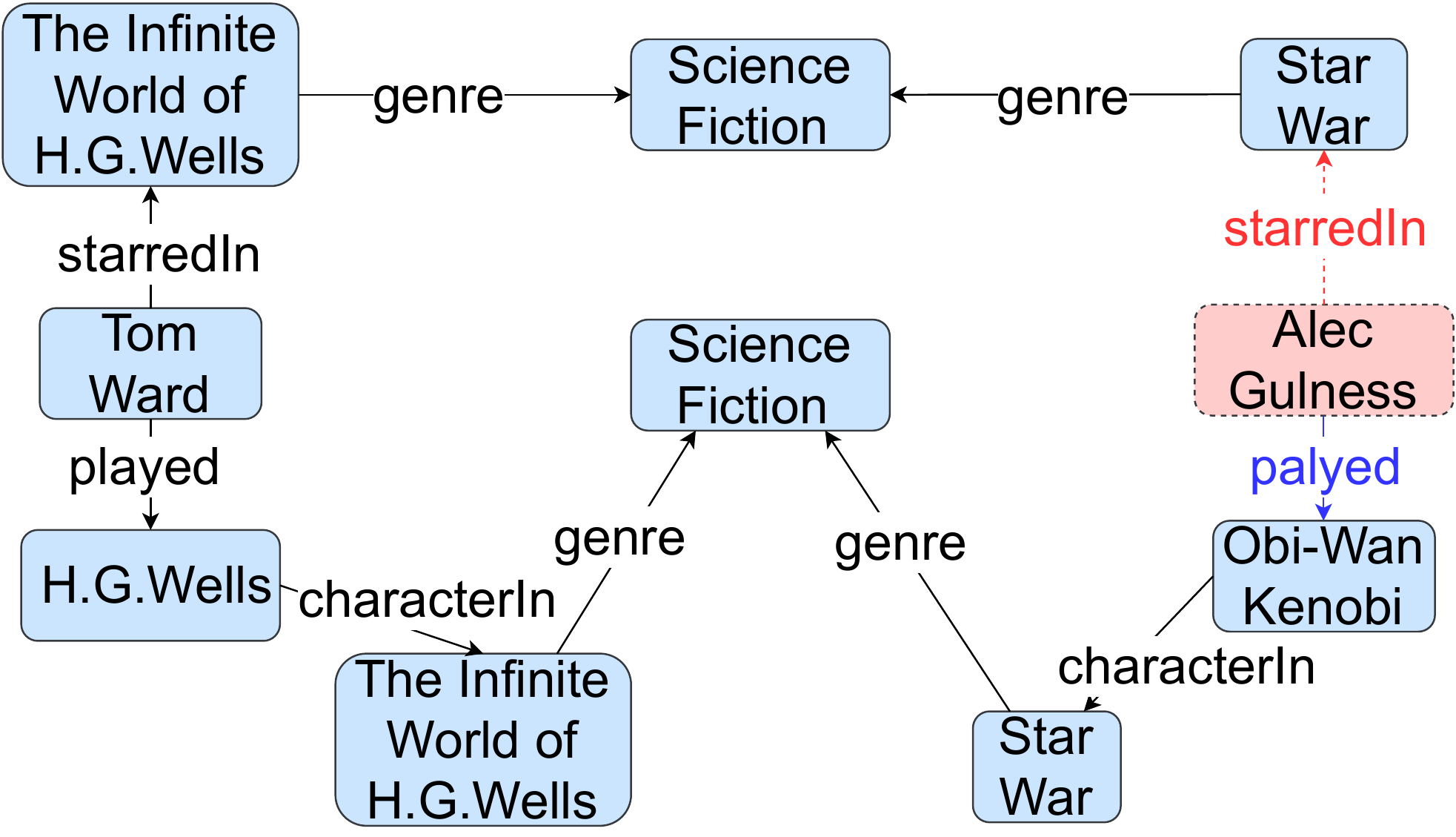}}
  \subfigure[Inter-rule correlation.]{
    \label{Fig: An example of the inter-rule correlation.}
    \includegraphics[width=0.48\linewidth]{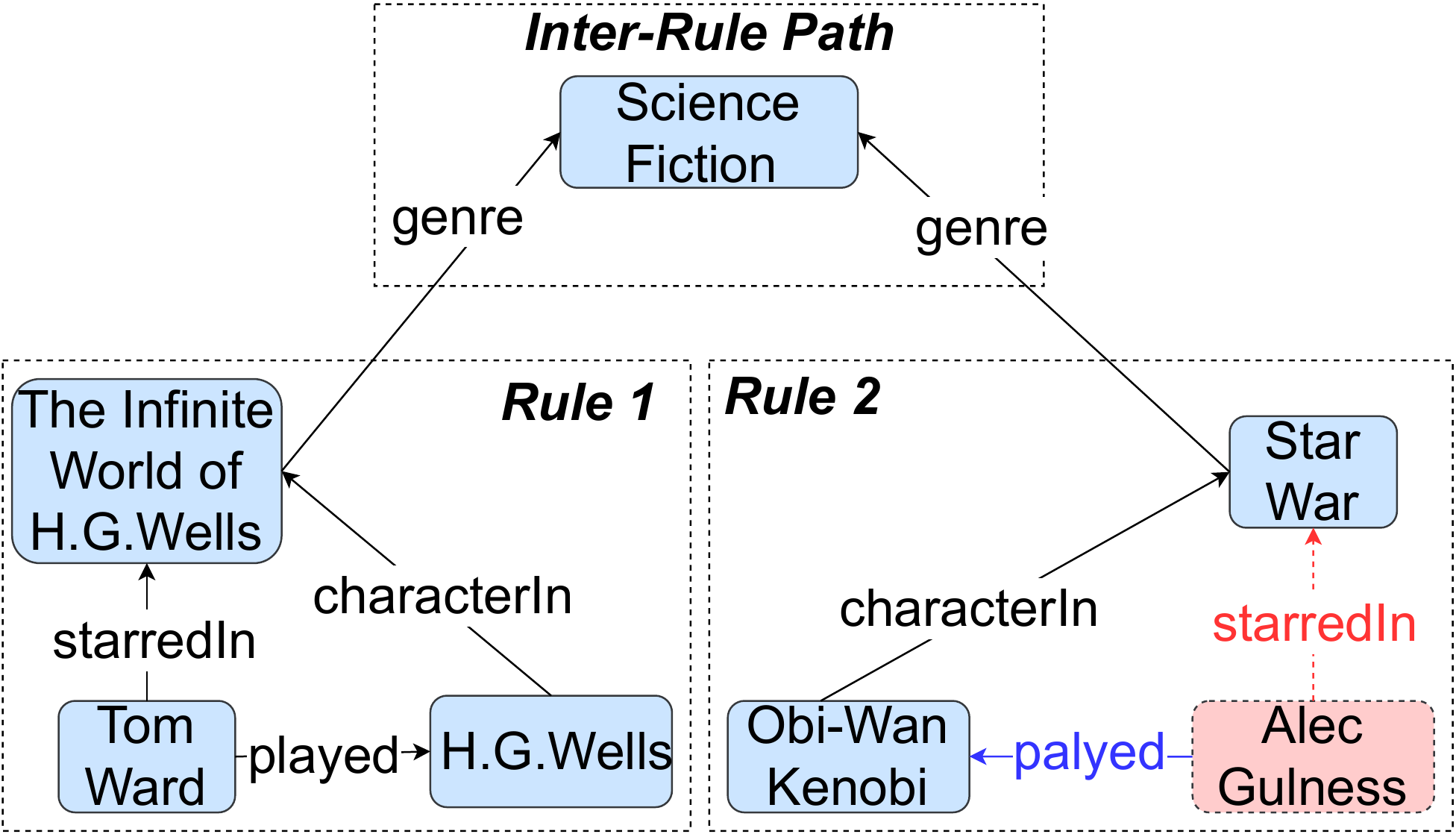}}
  \subfigure[FB15K Subject-20.]{
    \label{Fig: rule coverage on FB15K Subject-20}
    \includegraphics[width=0.48\linewidth]{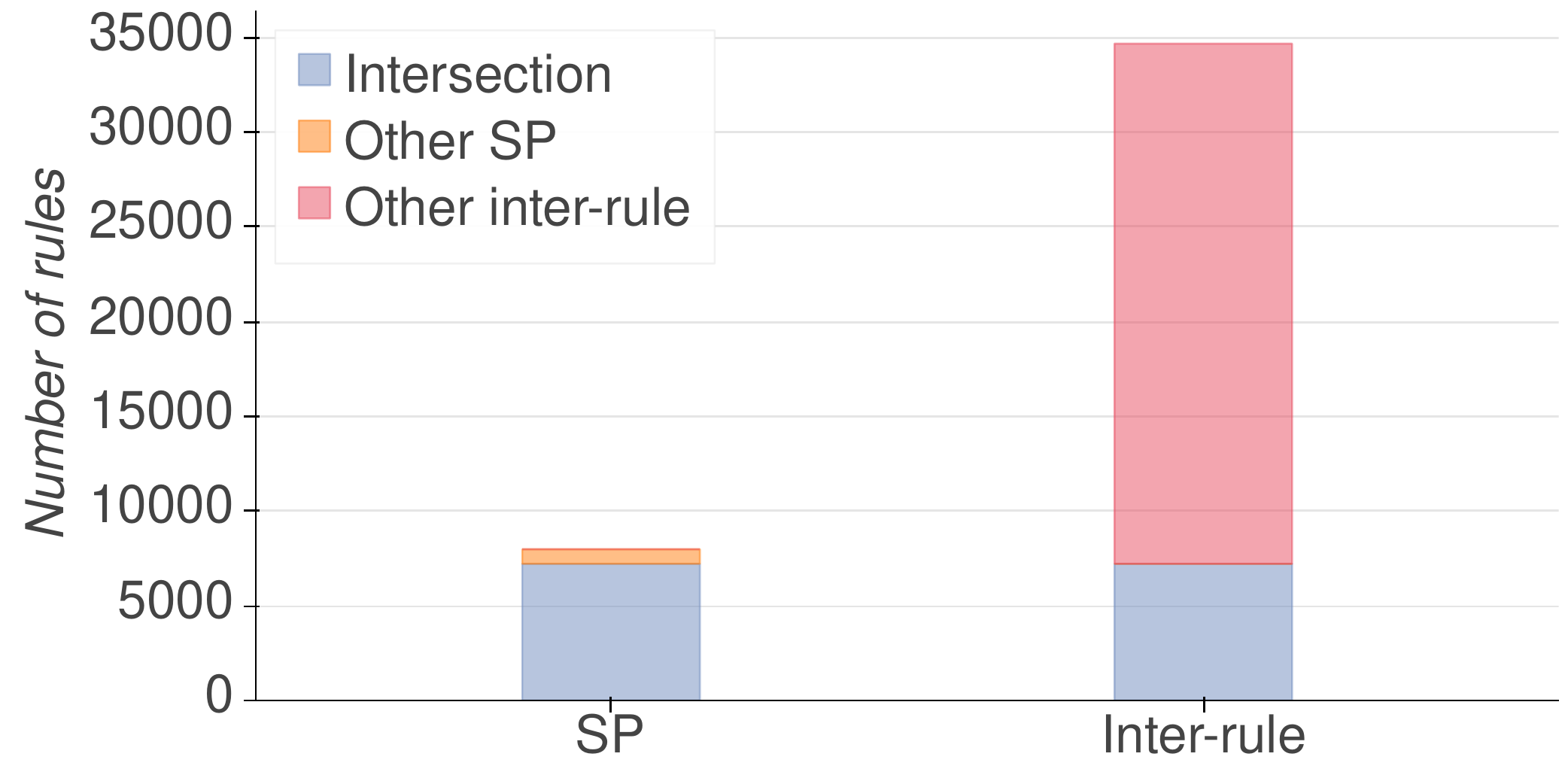}}
   \subfigure[WN18 Subject-20.]{
    \label{Fig: rule coverage on WN18 Subject-20}
    \includegraphics[width=0.48\linewidth]{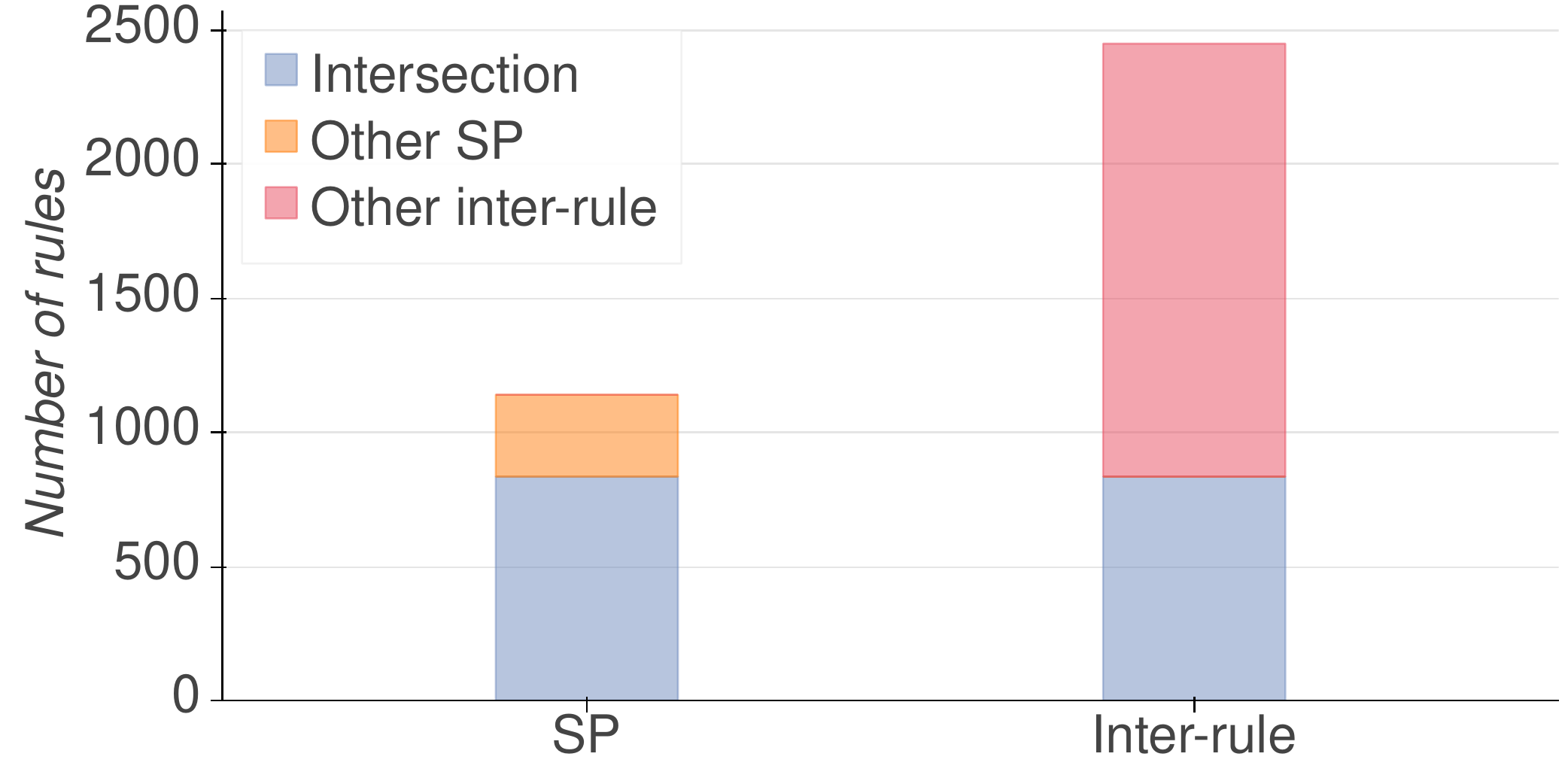}}
    \caption{(a) and (b) are examples of SP rules and inter-rule correlations, respectively; (c) and (d) demonstrate the intersection between SP rules and inter-rule correlations.}
  \label{Fig: symmetric path rule and inter-rule correlations}
\end{figure}

In VN network~\citep{DBLP:conf/cikm/HeWZTR20}, to capture long-distance semantic similarities between entities, symmetric path (SP) rules in \ac{KGs} are identified. 
In fact, many symmetric path rules can be transformed into inter-rule correlations.
For example, the SP rule shown in Fig.~\ref{Fig: An example of the symmetric path rule.} can be represented by the inter-rule correlation in Fig.~\ref{Fig: An example of the inter-rule correlation.}, since symmetric paths in the rule body and head share several entities and relations.
Motivated by this, as Fig.~\ref{Fig: rule coverage on FB15K Subject-20} and~\ref{Fig: rule coverage on WN18 Subject-20} shows, we count the number of the shared rules (blue bars) used by VN network and \ac{VNC} on FB15K Subject 20 and WN18 Subject 20.
The results indicate that the \ac{VNC} framework is capable of extracting most of the symmetric path rules (more than 80\%), and identifying abundant graph patterns to further alleviate the data sparsity problem, and improve the embedding quality.

\subsection{Rule inference}
\label{subsec: rule inference}
In the rule inference stage, given the extracted rules, our goal is to infer a new triple set with virtual neighbors $\mathcal{VN}$ and assign a soft label $s(t_{vn})$ to each predicted triple $t_{vn} \in \mathcal{VN}$.

\subsubsection{Rule modeling}
To predict a new triple $t_{vn}\in \mathcal{VN}$, we replace variables in extracted rules with concrete entities to obtain rule groundings.
To model rule groundings, we adopt t-norm based fuzzy logics \cite{DBLP:books/kl/Hajek98}. 
The key idea here is to compute the truth level of a grounding rule using the truth levels of its constituent triples and logical connectives (e.g., $\land$ and $\rightarrow$).
Following~\citep{DBLP:conf/aaai/GuoWWWG18,DBLP:books/kl/Hajek98}, logical connectives associated with the logical conjunction ($\land$), disjunction($\lor$), and negation ($\neg$) are defined as follows:
\begin{equation}
\small
\begin{split}
    I(a \land b) &= I(a) \cdot I(b),\\
    I(a \lor b) &= I(a) + I(b) - I(a) \cdot I(b),\\
    I(\neg a) &= 1 - I(a),
\end{split}
\label{eq:rules}
\end{equation}
where $a$ and $b$ denote logical expressions, which can be the atom triple or multiple triples combined by logical connectives. 
$I(\cdot)$ is the truth level function.
If $a = (e_{1},r_{1},e_{2})$ is a single triple, $I(a)$ is defined by Eq.~\ref{eq:DistMult}, i.e., $I(a) = \phi (e_{1},r_{1},e_{2})$.
For combined multiple triples, we can calculate the truth value using Eq.~\ref{eq:rules} recursively. 
For example, for a rule grounding $a \rightarrow b$, the truth value can be computed as: $I(a \rightarrow b) = I(\neg a \lor b) = I(a) \cdot I(b) - I(a) + 1$.
\subsubsection{Soft label prediction}
In this stage, our goal is to assign a soft label $s(x_{vn})\in [0,1]$ for each triple $t_{vn} \in \mathcal{VN}$,
using the current KG embeddings (i.e., $\mathbf{E}$ and $\mathbf{R}$) and rule groundings (i.e., $\mathcal{G}^{logic}$ and $\mathcal{G}^{corr}$).
To this end, we establish and solve a rule-constrained optimization problem.
Here, the optimal soft label $s(t_{\mathit{vn}})$ should keep close to truth level $I(t_{\mathit{vn}})$, while constrained by the rule groundings.
For the first characteristic, we minimize a square loss over the soft label $s(t_{vn})$ and truth level $I(t_{\mathit{vn}})$.
For the second characteristic, we impose rule constraints on the predicted soft labels $\mathcal{S}=\{s(t_{\mathit{vn}})\}$.
To be specific, given a rule $f_{m}$ and soft labels $\mathcal{S}$,  rule groundings $g_{\mathit{mn}}$ is expected to be true, i.e., $I(g_{\mathit{mn}}|\mathcal{S}) = 1$ with confidence $\lambda_{m}$. 
Here, the conditional truth level $I(g_{mn}|\mathcal{S})$ can be calculated recursively using the logical connectives in Eq.~\ref{eq:rules}.
Specifically, for each logic rule grounding $g^\mathit{logic}_{mn}: (e_{i}, r_{b}, e_{j}) \rightarrow (e_{i}, r_{h}, e_{j})$, where $(e_{i}, r_{b}, e_{j})\in \mathcal{O}$ and $(e_{i}, r_{h}, e_{j}) \in \mathcal{VN}$, the conditional truth level $I(g^\mathit{logic}_{mn}|\mathcal{S})$ is calculated as:
$
\label{Eq: conditional truth level for logic rules}
I(g^\mathit{logic}_{mn}|\mathcal{S}) = I(e_{i}, r_{b}, e_{j})\cdot s(e_{i}, r_{h}, e_{j}) - I(e_{i}, r_{b}, e_{j}) + 1,
$
where $I(e_{i}, r_{b}, e_{j})$ is the truth level defined in Eq.~\ref{eq:DistMult} computed using the current embedding, while $s(e_{i}, r_{h}, e_{j})$ is a soft label to infer.
Similarly, for each grounding of inter-rule correlations $g^{corr}_{vw}: g^\mathit{logic}_{b}\rightarrow g'{^\mathit{logic}}_{h}$, where $g^{logic}_{b}$ is a logic rule grounding and $g'{^\mathit{logic}}_{h}$ is a grounding for the corresponding ``incomplete'' logic rule, the conditional truth level $I(g^\mathit{corr}_{vw}|\mathcal{S})$ can be computed as:
\begin{equation}
\label{Eq: conditional truth level for inter-rule correlations}
    I(g^\mathit{corr}_{vw}|\mathcal{S}) = I(g^\mathit{logic}_{b})\cdot I(g'^{\mathit{logic}}_{h}|\mathcal{S}) - I(g^\mathit{logic}_{b}) + 1,
\end{equation}
where $I(g^{\mathit{logic}}_{b})$ is the truth level for the logic rule grounding $g^{\mathit{logic}}_{b}$, and $I(g'^{logic}_{h}|\mathcal{S})$ denotes the conditional truth level for the ``incomplete'' logic rule grounding $g'^{\mathit{logic}}_{h}$.
Since triples in $g^\mathit{logic}_{b}$ are observed in \ac{KG}, we suppose $I(g^\mathit{logic}_{b}) = 1$ for simplicity. 
Similar to Eq.~\ref{Eq: conditional truth level for logic rules}, the conditional truth level $I(g'^{\mathit{logic}}_{h}|\mathcal{S})$ can be computed recursively according to logical connectives.

Combining two characteristics, we introduce the slack variables $\xi^\mathit{logic}_{mn}$ and $\xi^\mathit{corr}_\mathit{vws}$, and establish the following optimization problem to obtain the optimal soft labels:
\begin{equation}
\small
    \begin{split}
    \min_{S, \xi}&{\frac{1}{2}\cdot \sum_{t_{vn} \in \mathcal{VN}}{(s(t_{vn})-I(t_{vn}))^2}}+C\cdot \left(\sum_{m,n}{\xi^\mathit{logic}_{mn}}+\sum_{v,w}{\xi^\mathit{corr}_{vw}}\right)
    \hspace*{-5mm}\mbox{}\\
    &\text{such that } \lambda^\mathit{logic}_{m}(1 - I(g^\mathit{logic}_{mn}|S)) \leq \xi^\mathit{logic}_{mn}\\
    &\phantom{\text{such that }}\lambda^\mathit{corr}_{v}(1 - I(g^\mathit{corr}_{vw}|S)) \leq \xi^\mathit{corr}_{vw} \\ 
    &\phantom{\text{such that }}\xi^\mathit{logic}_{mn} \geq 0, \xi^\mathit{corr}_{vw} \geq 0, 0\leq s(t_{vn})\leq 1,
    \end{split}
    \label{Eq: optimization problem}
\end{equation}
where $C$ is the constant penalty parameter, and $\lambda^\mathit{logic}_{m}$ and $\lambda^\mathit{corr}_{v}$ are the confidence values for logic rule $f^\mathit{logic}_{m}$ and inter-rule correlation $f^\mathit{corr}_{v}$ respectively.
Note that, for the optimization problem in Eq.~\ref{Eq: optimization problem}, all the constraints are linear functions w.r.t $s(t_{\mathit{vn}})$, and this kind of the optimization problem is convex \cite{DBLP:conf/aaai/GuoWWWG18}. 
Therefore, we can obtain the closed-form solution for this problem:
\begin{equation}
\small
\label{Eq: optimal soft labels}
\begin{split}
s(t_{vn}) = & \left[I(t_{\mathit{vn}}) + C\cdot \left( \sum_{m,n}{\lambda^{\mathit{logic}}_{m}\nabla_{s(t_{\mathit{vn}})}{I(g^{\mathit{logic}}_{mn}|S)}} \right. \right.\\
& \,\,\,\left. \left. +\sum_{v,w}{\lambda^{\mathit{corr}}_{v}\nabla_{s(t_{\mathit{vn}})}{I(g^{\mathit{corr}}_{vw}|S)}}\right)\right]_{0}^{1},\\
\end{split}
\end{equation}
where $\nabla_{s(t_{\mathit{vn}})}{I(g^{\mathit{logic}}_{mn}|S)}$ and $\nabla_{s(t_{\mathit{vn}})}{I(g^{\mathit{corr}}_{vw}|S)}$ denote the gradients of $I(g^{\mathit{logic}}_{mn}|S)$ and $I(g^{\mathit{corr}}_{vw}|S)$ w.r.t $s(t_{\mathit{vn}})$ respectively, which are both constants. 
$[\cdot]_{0}^{1} = \min(\max(x, 0), 1)$ is a truncation function. 

\subsection{Embedding}
\label{subsec: embedding}
In the embedding stage, the knowledge graph with softly labeled virtual neighbors is inputted into the GNN-based encoder and embedding-based decoder.
In this way, entities and relations in \ac{KG} are projected into embeddings $\mathbf{E}$ and $\mathbf{R}$.

\subsubsection{GNN-based encoder}
Similar to previous works~\citep{DBLP:conf/aaai/WangHLP19,DBLP:conf/cikm/HeWZTR20},
our encoder consists of several \emph{structure aware} layers and one \emph{query aware} layer.
To model connectivity structures of the given \ac{KG}, we adopt weighted graph convolutional network (WGCN) \cite{DBLP:conf/aaai/ShangTHBHZ19} as the structure aware layers.
In each layer, different relation types are assigned distinct attention weights. 
The $l$-th structure aware layer can be formulated as follows:
\begin{equation}
\small
\label{Eq: WGCN}
\begin{split}
\mathbf{a}_{i}^{(l)} &=  \mathbf{W}^{(l)}\cdot\left(\sum_{(e_{i},r,e_{j}) \in \mathcal{O}\cup \mathcal{VN}}{\alpha_{r}^{(l)}\mathbf{h}^{(l-1)}_{j}}\right),\\
\mathbf{h}_{i}^{(l)} &=  \tanh\left(\mathbf{a}_{i}^{(l)} + \mathbf{h}_{i}^{(l-1)}\mathbf{W}^{(l)}\right),
\end{split}
\end{equation}
where $\alpha_{r}$ are the attention weights for relation $r$.
$\mathbf{h}_{i}^{(l)}$ is the embedding of entity $e_{i}$ at the $l$ th layer. 
$\mathbf{W}^{(l)}$ is the connection matrix for the $l$ th layer.
Here, we randomly initialize the input entity embedding $\mathbf{h}^{(0)}_{i}$ during training. 
Besides the structure information, given the query relation in each inputted triple, an ideal aggregator is able to focus on the relevant facts in the neighborhood.
To this end, the importances of neighbors are calculated based on the neural network mechanism \cite{DBLP:conf/aaai/WangHLP19}. 
Specifically, given a query relation $r_{q}\in \mathcal{R}$, the importance of the neighbor $e_{j}$ to entity $e_{i}$ is calculated as:
$
\alpha^{\rm NN}_{j|i, q} = {\rm softmax}(\beta_{j|i,q}) = \frac{{\rm exp}(\beta_{j|i,q})}{\sum_{(e_{i},r_{q},e_{j'})\in \mathcal{O}\cup\mathcal{VN}}{{\rm exp}(\beta_{j'|i,q})}},
$
where the unnormalized attention weight $\beta_{j|i,q}$ can be computed as: $\beta_{j|i, q} = {\rm LeakyReLU}(\mathbf{u}\cdot[\mathbf{W}_{e}\mathbf{h}_{i};\mathbf{W}_{q}\mathbf{z}_{q};\mathbf{W}_{e}\mathbf{h}_{j}])$,
where $\mathbf{u}$, $\mathbf{W}_{e}$, and $\mathbf{W}_{q}$ are the attention parameters, 
and $\mathbf{z}_{q}$ is the relation-specific parameter for query relation $r_{q}$.
${\rm LeakyReLU}(\cdot)$ is the activation function of the leaky rectified linear unit~\citep{DBLP:journals/corr/XuWCL15}.
On this basis, we can formulate the query aware layer as follows:
\begin{equation}
    \label{Eq: query aware layer}
  \mathbf{h}^{O}_{i}= \sum_{(e_{i},r,e_{j})\in \mathcal{O}\cup\mathcal{VN}}{\alpha^{\rm NN}_{j|i, q} \cdot \mathbf{h}^{I}_{j}},
\end{equation}
where $\mathbf{h}^{I}_{j}$ is the input embedding for the entity $e_{j}$ from the last structure aware layer.
$\mathbf{h}^{O}_{i}$ is the output embedding for the entity $e_{i}$ for the decoder.
Note that, in the testing process, we apply the encoder on the auxiliary triples, 
and initialize the input representation $\mathbf{h}^{(0)}_{i'}$ for the \ac{OOKG} entity $e_{i'}$ as the zero vector.

\subsubsection{Embedding-based decoder}
Given entity embeddings from the GNN-based encoder (i.e., $\mathbf{E}=\mathbf{H}^{O}$, where $\mathbf{H}^{O}$ is the output of the encoder), the decoder aims to learn relation embeddings $\mathbf{R}$, and compute the truth level $\phi(t)$ for each triple $t$.
We evaluate various embedding methods in our experiments, including DistMult \cite{DBLP:journals/corr/YangYHGD14a}, TransE~\citep{DBLP:conf/nips/BordesUGWY13}, ConvE~\citep{DBLP:conf/aaai/DettmersMS018}, and  Analogy~\citep{DBLP:conf/icml/LiuWY17} (see \S\ref{sec:Influence of decoders}).
\subsection{Training algorithm}
\label{subsec: training algorithm}
To refine the current KG embeddings, a global loss over facts with hard and soft labels is utilized in the \ac{VNC} framework.
In this stage, we randomly corrupt the head or tail entity of an observed triple to form a negative triple.
In this way, in addition to triples with soft labels $\mathcal{VN} = \{t_{vn}\}$, we collect the observed and negative fact triples with hard labels, i.e., $\mathcal{L}=\{x_{l}, y_{l}\}$, where $y_{l} \in \{0, 1\}$ is the hard label of the triples.
To learn the optimal KG embeddings $\mathbf{E}$ and $\mathbf{R}$, a global loss function over $\mathcal{L}$ and $\mathcal{VN}$ is:
\begin{equation}
\small
\min_{\mathbf{E}, \mathbf{R}}{\frac{1}{|\mathcal{L}|}\sum_{\mathcal{L}}{l(I(t_{l}), y_{l})} + \frac{1}{|\mathcal{VN}|}\sum_{\mathcal{VN}}{l(I(t_{vn}), s(t_{vn}))}},
\label{Eq: loss}
\end{equation}
where we adopt the cross entropy $l(x,y) = -y\log{x} - (1-y)\log{(1-x)}$.
$I(\cdot)$ is the truth level function.
We use Adam \cite{DBLP:journals/corr/KingmaB14} to minimize the global loss function.
In this case, the resultant KG embeddings fit the observed facts while constrained by rules.
Algorithm~\ref{Algorithm: training algorithm of VNC} summarizes the training process of \ac{VNC}.
Before training, rule pools are generated by finding plausible paths, and rules of low quality are filtered out (line 1). 
In each training step, we compute rules $\lambda^{\mathit{logic}}$ and $\lambda^{\mathit{corr}}$ using current relation embeddings to form rule sets $\mathcal{F}^{\mathit{logic}}$ and  $\mathcal{F}^{\mathit{corr}}$ (line 3).
Then, in the rule inference stage, we infer new triples $\mathcal{VN} = \{t_{\mathit{vn}}\}$ using rule groundings, and assign a soft label $s{t_{\mathit{vn}}}$ to each predicted fact triples by solving a rule constrained optimization problem (line 4-6).
Next, the knowledge graph with virtual neighbors is inputted into the GNN-based encoder and embedding-based decoder.
In this way, relations and entities are mapped into embeddings (line 7).
Finally, the overall loss over fact triples with hard and soft labels is obtained (line 8-9), and embeddings as well as model parameters are updated (line 10). 


\begin{algorithm}[t]
\small
    \caption{Training algorithm of \ac{VNC}.}
    \label{Algorithm: training algorithm of VNC}      
    \begin{algorithmic}[1] 
    \Require Triples with hard labels $\mathcal{L}$; randomly initialized entity and relation embeddings $\mathbf{E}$ and $\mathbf{R}$; parameters $\mathbf{\Theta}$ for encoder and decoder.
    \Ensure The extracted rule set $\mathcal{F}^{\mathit{logic}}$ and  $\mathcal{F}^{\mathit{corr}}$; trained encoder and decoder; optimal embeddings $\mathbf{E}$ and $\mathbf{R}$;
    \State Generate rule pools, and filter out rules of low quality;
    
    \While{Training process not terminated}
    \State Compute rule confidences $\lambda^{\mathit{logic}}$ and $\lambda^{\mathit{corr}}$, and form rule sets $\mathcal{F}^{\mathit{logic}}$ and  $\mathcal{F}^{\mathit{corr}}$ (Eq.~\ref{Eq: similarity funciton} and~\ref{Eq: inter-rule confidence});
    
    \State Find rule groundings $\mathcal{G}^{\mathit{logic}}$ and $\mathcal{G}^{\mathit{corr}}$;
    
    \State Infer $\mathcal{VN} = \{t_{\mathit{vn}}\}$ and compute the truth level $I(t_{\mathit{vn}})$ for each triple $t_{\mathit{vn}}$ (Eq.~\ref{eq:DistMult});
    
    \State Calculate the conditional truth level $I(g|\mathcal{S})$ (Eq.~\ref{Eq: conditional truth level for inter-rule correlations});
    
    \State Obtain the optimal soft labels $\mathcal{S} = \{s(t_{\mathit{vn}})\}$ (Eq.~\ref{Eq: optimization problem} and~\ref{Eq: optimal soft labels});
    
    \State Obtain embeddings $\mathbf{E}$ and $\mathbf{R}$ (Eq.~\ref{eq:DistMult}, ~\ref{Eq: WGCN}, and ~\ref{Eq: query aware layer});
    
    \State Compute the global loss over $\mathcal{L}$ and $\mathcal{VN}$ (Eq.~\ref{Eq: loss});
    
    \State Update $\mathbf{E}$, $\mathbf{R}$ and $\mathbf{\Theta}$;
 \EndWhile
    \end{algorithmic}
    \end{algorithm}


\section{Experiments}
\label{sec:Experiments}
\textbf{Research questions.}
We aim to answer the following research questions:
\begin{enumerate*}[label=(RQ\arabic*)]
    \item 
    Does \ac{VNC} outperform state-of-the-art methods on the OOKG entity problem? (See \S\ref{sec:link prediction}--\S\ref{sec:Comparisions with entity-independent methods}.)
    \item How do the inter-rule correlations and iterative framework contribute to the performance? (See \S\ref{sec:Ablation studies}.)
    \item What is the influence of the decoder, embedding dimension, and penalty parameter on the performance? (See \S\ref{sec:Influence of decoders}.)
    \item Is \ac{VNC} able to identify useful inter-rule correlations in the knowledge graph? (See \S\ref{sec:Case study}.)
\end{enumerate*}

\header{Datasets}
\label{subsec:datasets}
We evaluate \ac{VNC} on three widely used datasets: YAGO37~\citep{DBLP:conf/aaai/GuoWWWG18}, FB15K~\citep{DBLP:conf/nips/BordesUGWY13}, and WN11~\citep{DBLP:conf/nips/SocherCMN13}.
For link prediction, we use three benchmark datasets: YAGO37 and FB15K. We create Subject-R and Object-R from each benchmark dataset, varying \ac{OOKG} entities' proportion ($R$) as 5\%, 10\%, 15\%, 20\%, and 25\% following~\citep{DBLP:conf/aaai/WangHLP19}.
For triple classification, we directly use the datasets released in~\citep{DBLP:conf/ijcai/HamaguchiOSM17} based on WN11, including Head-$N$, Tail-$N$ and Both-$N$, 
where $N = \{1000, 3000, 5000\}$ testing triples are randomly sampled to construct new datasets.
Tab.~\ref{Tab:Statistics of datasets} gives detailed statistics of the datasets.

\header{Baselines}
We compare the performance of \ac{VNC} against the following baselines:
\begin{enumerate*}[label=(\roman*)]
    \item \textbf{MEAN}~\citep{DBLP:conf/ijcai/HamaguchiOSM17} utilizes the \acf{GNN} and generates embeddings of \ac{OOKG} entities with simple pooling functions.
    \item \textbf{LSTM}~\citep{DBLP:conf/aaai/WangHLP19} is a simple extension of MEAN, where the LSTM network~\citep{DBLP:conf/nips/HamiltonYL17} is used due to its large expressive capability.
    \item \textbf{LAN}~\citep{DBLP:conf/aaai/WangHLP19} uses a logic attention network as the aggregator to capture information of redundancy and query relations in the neighborhood.
    \item \textbf{GEN}~\citep{DBLP:conf/nips/BaekLH20} develops a meta-learning framewrok to simulate the unseen entities during meta-training.
    \item \textbf{VN network}~\citep{DBLP:conf/cikm/HeWZTR20} infers additional virtual neighbors for \ac{OOKG} entities to alleviate the data sparsity problem.
    \item \textbf{InvTransE} and \textbf{InvRotatE}~\citep{DBLP:journals/corr/abs-2009-12765} obtain optimal estimations of \ac{OOKG} entity embeddings with translational assumptions.
\end{enumerate*}
Besides, we also compare \ac{VNC} with the following entity-independent embedding methods. 
\begin{enumerate*}[label=(\roman*)]
    \item \textbf{DRUM}~\citep{DBLP:conf/nips/SadeghianADW19} designs an end-to-end rule mining framework via the connection between tensor completion and the estimation of confidence scores.
    \item \textbf{GraIL}~\citep{DBLP:conf/icml/TeruDH20} is a \ac{GNN} framework that reasons over local subgraphs and learns entity-independent relational semantics.
    \item \textbf{TACT}~\citep{DBLP:conf/aaai/ChenHWW21} incorporates seven types of semantic correlations between relations with the existing inductive methods.
\end{enumerate*}
For entity-independent methods, the training sets are considered as the original \ac{KG}s while training sets with auxiliary facts are regarded as the new \ac{KG}s during testing.

\header{Evaluation metrics}
For the link prediction task, we report filtered Mean Rank (MR), Mean Reciprocal Rank (MRR), and Hits at $n$ (Hits@$n$), where filtered metrics are computed after removing all the other positive triples that appear in either training, validation, or test set during ranking. For the triple classification task, models are measured by classifying a fact triple as true or false, and Accuracy is applied to assess the proportion of correct triple classifications.

\header{Implementation details}
We fine-tune hyper-parameters based on validation performance. Encoders and decoders have 200 dimensions. Learning rate, dropout, and regularization are set to 0.02, 0.2, and 0.01. $C$ is maintained at 1. The GNN-based encoder comprises two structure-aware layers and one query-aware layer, while DistMult serves as the decoder. For WN18, $\alpha_{HC}$ and $\alpha_{SC}$ are 0.3; for FB15K, they are 0.5. Optimal values for $\alpha_{HC}$ and $\alpha_{SC}$ on YAGO37 and WN11 are 0.01. Across all datasets, $\alpha_{PCRA}$ is set to 0.01.

\begin{table}
    \caption{Descriptive statistics of the datasets.}
     \label{Tab:Statistics of datasets}
     \resizebox{\linewidth}{!}{
        \begin{tabular}{l rr rrr}  
        \toprule
        Dataset & Entities & Relations & Training & Validation & Test\\
        \midrule
        YAGO37 & 123,189 & 37 & 989,132 & 50,000 & 50,000\\
        FB15K & 14,951 & 1,345 & 483,142 & 50,000 & 59,071\\
        WN11 & 38,696 & 11 & 112,581 & 2,609 & 10,544\\
        \bottomrule
        \end{tabular}
        }
\end{table}

\section{Results}
\label{sec:Results}

\subsection{Link prediction (RQ1)}
\label{sec:link prediction}
\begin{figure}
  \centering
    \includegraphics[width=0.95\linewidth]{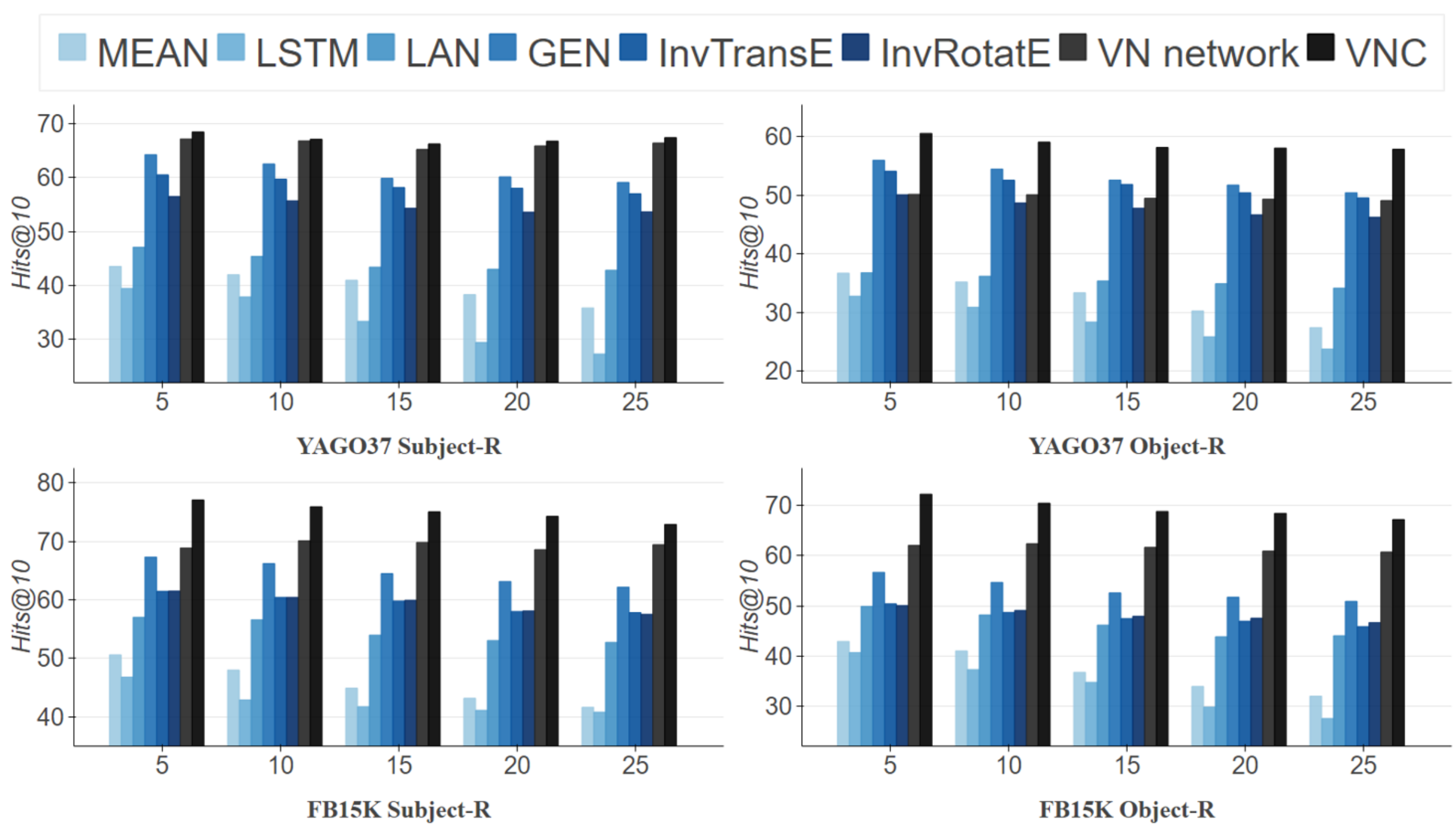}
    \caption{Link prediction results on YAGO37 and FB15K.
    }
  \label{Fig: Evaluation results (Hits@10) of link prediction on YAGO37 and FB15K.}
\end{figure}

Tab.~\ref{Tab:Evaluation results on YAGO37} and Fig.~\ref{Fig: Evaluation results (Hits@10) of link prediction on YAGO37 and FB15K.} show the experimental outcomes for the link prediction task.
Based on the experimental results, we have the following observations:
\begin{enumerate*}[leftmargin=*, nosep, label=(\roman*)]
    \item Link predictions for \ac{OOKG} entities are challenging, and for most baselines, the Hits@n and MRR are less than 0.7. In contrast, the proposed \ac{VNC} is able to effectively infer missing fact triples for unseen entities.
    \item The proposed model \ac{VNC} consistently outperforms state-of-the-art baselines and VN network over all the datasets. 
    Compared to the baseline models, \ac{VNC} achieves considerable increases in all metrics, including MR, MRR, Hits@10, Hits@3, and Hits@1. 
    That is, identifying inter-rule correlations and capturing interactions among rule mining, rule inference, and embedding substantially enhance the performance for \ac{OOKG} entities. 
    \item When the ratio of the unseen entities increases and observed \ac{KG}s become sparser, \ac{VNC} is still able to accurately predict missing triples for \ac{OOKG} entities.
    In Fig.~\ref{Fig: Evaluation results (Hits@10) of link prediction on YAGO37 and FB15K.}, we show the results of link prediction experiments on datasets with different sample rates $R$. 
    As the number of unseen entities increases, \ac{VNC} still maintains the highest Hits@10 scores, indicating its robustness on sparse \acp{KG}.
\end{enumerate*}
In summary, recognizing inter-rule correlations in \ac{KG}s and designing an iterative framework for rule and embedding learning is able to strengthen the performance.

\subsection{Triple classification (RQ1)}
\label{sec:triple classification}
To further evaluate \ac{VNC}, we conduct triple classification on the WN11 dataset. 
Based on the evaluation results in Tab.~\ref{Tab:Evaluation results (Accuracy) of triple classification on WN11}, we observe that \ac{VNC} achieves state-of-the-art results on the triple classification task.
With shallow pooling functions, MEAN and LSTM lead to the lowest accuracy. 
Meanwhile, other baseline models are hindered by the data sparsity problem, and ignoring complex patterns in graphs and interactions between rule and embedding learning.
In contrast, \ac{VNC} infers virtual neighbors for \ac{OOKG} entities and mine logic rules and inter-rule correlations from \ac{KG}s in an iterative manner, which results in the highest accuracies over all the datasets.

\subsection{Comparisons with entity-independent methods (RQ1)}
\label{sec:Comparisions with entity-independent methods}

In addition to the entity-specific baselines, we compare \ac{VNC} against entity-independent methods . 
Tab.~\ref{Tab:Comparisions with entity-independent methods on FB15K and WN18} shows the evaluation results on FB15K Subject-10. We draw the following conclusions:
\begin{enumerate*}[leftmargin=*, nosep, label=(\roman*)]
    \item In comparison with entity-independent methods, the state-of-the-art entity-specific frameworks perform better, demonstrating the importance of embeddings of known entities.
    Compared to DRUM, GraIL, and TACT, entity-specific embedding models, including GEN, InvTransE, InvRotatE, VN network, and \ac{VNC}, utilize pretrained embeddings of observed entities and attain huge performance enhancements.
    \item \ac{VNC} outperforms both entity-independent and entity-specific embedding methods, and achieves the best performance. This is, for \ac{OOKG} entities, identifying inter-rule correlations in \acp{KG} and aggregating embeddings of neighborhood entities facilitate predictions of missing facts.
\end{enumerate*}
In summary, extracting inter-rule correlations iteratively and integrating with embeddings of observed entities benefits the \ac{OOKG} entity problem.
\begin{table*}
\caption{
Link prediction results on YAGO37 and FB15K.
Significant improvements over the best baseline are marked with $\ast$ (t-test, $p < 0.05$).}
\label{Tab:Evaluation results on YAGO37}
\setlength{\tabcolsep}{0.5mm}
\resizebox{\linewidth}{!}{%
\begin{tabular}{@{} l @{~} ccccc ccccc ccccc ccccc @{}}
\toprule
		&\multicolumn{10}{c}{YAGO37}&\multicolumn{10}{c}{FB15K}\\
	\cmidrule(r){2-11} \cmidrule(r){12-21}
        &\multicolumn{5}{c}{Subject-10}&\multicolumn{5}{c}{Object-10} & \multicolumn{5}{c}{Subject-10}&\multicolumn{5}{c}{Object-10} \\
	\cmidrule(r){2-6} \cmidrule(r){7-11} \cmidrule(r){12-16} \cmidrule{17-21}
    Model &MR &MRR &Hits@10 &Hits@3 & Hits@1&MR &MRR &Hits@10 &Hits@3 & Hits@1 &MR &MRR &Hits@10 &Hits@3 & Hits@1&MR &MRR &Hits@10 &Hits@3 & Hits@1 \\
    \midrule
    MEAN & 2393 &21.5 &42.0 &24.2 &17.8 &4763 &17.8 &35.2 &17.5 &12.1 & 293 & 31.0 & 48.0 & 34.8 & 22.2 & 353 & 25.1 & 41.0 & 28.0 & 17.1 \\
    LSTM & 3148 &19.4 &37.9 &20.3 &15.9 &5031 &14.2 &30.9 &16.1 &11.8 & 353 & 25.4 & 42.9 & 29.6 & 16.2 & 504 & 21.9 & 37.3 & 24.6 & 14.3 \\
    LAN & 1929 &24.7 &45.4 &26.2 &19.4 &4372 &19.7 &36.2 &19.3 &13.2 & 263 & 39.4 & 56.6 & 44.6 & 30.2 & 461 & 31.4 & 48.2 & 35.7 & 22.7  \\
    GEN & 2259 &46.9 &62.5 &52.5 &39.3 &4258 &36.2 &54.5	&42.3	&27.7 & 165 & 47.5 & 66.2 & 54.3 & 38.7 & 201 & 44.1 & 54.7 & 43.8 & 31.8\\
    InvTransE & 2308 &44.9 &59.7 &50.2 &37.7 &4438 &35.0 &52.6 &40.9 &26.8 & 218 & 46.2 & 60.4 & 50.3 & 38.5 & 315 & 35.7 & 48.7 & 38.4 & 29.0\\
    InvRotatE &2381 &42.1 &55.7 &47.1 &35.3 &4518 &32.5 &48.7 &37.6 &24.9 & 233	& 45.3 & 60.4 & 50.2 & 36.9 & 276 & 36.2 & 49.1 & 38.6 & 29.3\\
    VN network & 1757 & 46.5 & 66.8 & 53.8 & 35.7 & 3145 & 27.4 & 50.1 & 36.4 & 19.5 & 175 & 46.3 & 70.1 & 52.6 & 34.5 & 212 & 42.3 & 62.7 & 44.6 & 28.2\\
    \midrule
    VNC &\textbf{1425}\rlap{$^{\ast}$} &\textbf{50.6}\rlap{$^{\ast}$} &\textbf{67.1} &\textbf{56.5}\rlap{$^{\ast}$} &\textbf{42.3}\rlap{$^{\ast}$} &\textbf{2638}\rlap{$^{\ast}$} &\textbf{39.1}\rlap{$^{\ast}$} &\textbf{59.1}\rlap{$^{\ast}$} &\textbf{45.8}\rlap{$^{\ast}$} &\textbf{30.1}\rlap{$^{\ast}$} & \textbf{151}\rlap{$^{\ast}$} & \textbf{54.3}\rlap{$^{\ast}$} & \textbf{75.9}\rlap{$^{\ast}$} & \textbf{60.8}\rlap{$^{\ast}$} & \textbf{41.6}\rlap{$^{\ast}$} & \textbf{183}\rlap{$^{\ast}$} & \textbf{48.2}\rlap{$^{\ast}$} & \textbf{70.5}\rlap{$^{\ast}$} & \textbf{52.9}\rlap{$^{\ast}$} & \textbf{36.3}\rlap{$^{\ast}$}\\
    \bottomrule
    \end{tabular}
    }
\end{table*}
\begin{table}
    \caption{
Triple classification results on WN11.
Significant improvements over the best baseline are marked with $\ast$ (t-test, $p < 0.05$).}
 \label{Tab:Evaluation results (Accuracy) of triple classification on WN11}
 \setlength{\tabcolsep}{1mm}
\resizebox{\linewidth}{!}{%
    \begin{tabular}{l @{~} ccc ccc ccc}  
    \toprule
          &\multicolumn{3}{c}{Subject}&\multicolumn{3}{c}{Object}&\multicolumn{3}{c}{Both}\\
    \cmidrule(r){2-4}
    \cmidrule(r){5-7}
    \cmidrule{8-10}
    Model &1000 &3000 &5000 &1000 & 3000& 5000 & 1000 & 3000 & 5000\\        
    \midrule
    MEAN &87.3 & 84.3 & 83.3 & 84.0 & 75.2 & 69.2 & 83.0 & 73.3 & 68.2\\
    LSTM &87.0 & 83.5 & 81.8 & 82.9 & 71.4 & 63.1 & 78.5 & 71.6 & 65.8\\
    LAN & 88.8 & 85.2 & 84.2 & 84.7 & 78.8 & 74.3 & 83.3 & 76.9 & 70.6\\
    GEN & 88.6 & 85.1 & 84.6 & 84.1	& 77.9 & 74.4 & 85.1 & 76.2 & 73.9\\
    InvTransE & 88.2 & 87.8 & 83.2 & 84.4 & 80.1 & 74.4 & 86.3 & 78.4 & 74.6\\
    InvRotatE & 88.4 & 86.9 & 84.1 & 84.6 & 80.1 & 74.9 & 84.2 & 75.0 & 70.6\\
    VN network & 89.1 & 85.9 & 85.4 & 85.5 & 80.6 & 76.8 &84.1 &78.5 &73.1\\
    \midrule
    VNC & \textbf{90.6}\rlap{$^{\ast}$} & \textbf{88.9}\rlap{$^{\ast}$} & \textbf{86.7}\rlap{$^{\ast}$} & \textbf{86.9}\rlap{$^{\ast}$} & \textbf{82.3}\rlap{$^{\ast}$} & \textbf{78.3}\rlap{$^{\ast}$} & \textbf{87.7}\rlap{$^{\ast}$} & \textbf{79.6}\rlap{$^{\ast}$} & \textbf{76.2}\rlap{$^{\ast}$}\\
    \bottomrule
    \end{tabular}
    }
\end{table}
\begin{table}
\caption{
Link prediction results on FB15K Subject-10.
Significant improvements over the best baseline are marked with $\ast$ (t-test, $p < 0.05$).}
\label{Tab:Comparisions with entity-independent methods on FB15K and WN18}
\setlength{\tabcolsep}{1.5mm}
\begin{tabular}{@{} l @{~} ccccc ccccc @{}}
\toprule
    Model &MR &MRR &Hits@10 &Hits@3 & Hits@1 \\
    \midrule
    DRUM & 249 & 41.6 & 59.4 & 46.8 & 31.7 \\
    GraIL & 241 & 41.9 & 60.1 & 47.3 & 32.1 \\
    TACT & 238 & 42.6 & 60.2 & 47.1 & 32.9 \\
    \midrule
    VNC & \textbf{151}\rlap{$^{\ast}$} & \textbf{54.3}\rlap{$^{\ast}$} & \textbf{75.9}\rlap{$^{\ast}$} & \textbf{60.8}\rlap{$^{\ast}$} & \textbf{41.6}\rlap{$^{\ast}$} \\
    \bottomrule
    \end{tabular}
\end{table}

\section{Analysis}
\label{sec:Analysis}

\subsection{Ablation studies (RQ2)}
\label{sec:Ablation studies}

To evaluate the effectiveness of each component in the \ac{VNC} framework, we conduct ablation studies on the link prediction task. 
The results are shown in Tab.~\ref{Tab:Ablation studies on FB15K Subject-10}. 
When only employing the GNN-based encoder and embedding-based decoder (``no rules''), all metrics suffer a severe drop.
In the ``hard rules'' setting, virtual neighbors are directly inferred by logic rules instead of soft label predictions. 
Compared to the ``no rules'' settings,
predicting virtual neighbors with hard logic rules effectively alleviates the data sparsity problem. 
To examine the necessity of the iterative framework, we extract logic rules and learn knowledge embeddings simultaneously in the ``soft rules'' setting.
The results show that the iterative framework captures interactions among rule mining, rule inference, and embedding, and gains considerable improvements over the model with hard logic rules. 
Moreover, compared with the ``soft rules'' setting, \ac{VNC} further improves the performance by identifying inter-rule correlations in \ac{KG}.
In short, both inter-rule correlations and the iterative framework contribute to the improvements in performance.
We also consider two model variants, \ac{VNC} (AMIE+) and \ac{VNC} (IterE), with different rule mining frameworks AMIE+~\citep{DBLP:journals/vldb/GalarragaTHS15} and IterE~\citep{DBLP:conf/www/ZhangPWCZZBC19}, respectively. 
\ac{VNC} (AMIE+) mines logic rules with AMIE+, and keeps confidence scores of logic rules unchanged during the training process. 
\ac{VNC} (IterE) assumes the truth values of triples existing in \acp{KG} to be 1, and then calculates soft labels recursively using Eq.~\ref{eq:rules} instead of solving the optimization problem in Eq.~\ref{Eq: optimization problem}.
The results in Tab.~\ref{Tab:Ablation studies on FB15K Subject-10} show that the proposed iterative framework in \ac{VNC} outperforms other rule mining methods, indicating the effectiveness of \ac{VNC}.

\subsection{Influence of decoder (RQ3)}
\label{sec:Influence of decoders}

To assess the impact of various decoders on performance, we examine four types of embedding-based decoders, including TransE~\citep{DBLP:conf/nips/BordesUGWY13}, ConvE~\citep{DBLP:conf/aaai/DettmersMS018}, Analogy~\citep{DBLP:conf/icml/LiuWY17}, and DistMult~\citep{DBLP:journals/corr/YangYHGD14a}, regarding their effectiveness in the link prediction task.
According to the results in Tab. \ref{Tab:Influences of different decoders on FB15K Subject-10}, \ac{VNC} using the TransE decoder demonstrates the lowest performance, while \ac{VNC} with DistMult achieves the highest performance. In comparison to translational models, the bilinear scoring function-based decoder is more compatible with our framework.

\subsection{Case studies (RQ4)}
\label{sec:Case study}
For RQ4, we conduct case studies on \ac{VNC}, and Tab.~\ref{Tab: case study} shows examples of the inter-rule correlations on YAGO37.
In the first example, from the logic rule and inter-rule path, it is easy to find that ``George'' is the director and producer of ``Young Bess'' and ``Cass Timberlane''.
Similarly, the second example shows that the children usually have the same citizenship as their parents.
Note that, the above missing facts can not be directly inferred by either logic rules or symmetric path rules~\citep{DBLP:conf/cikm/HeWZTR20}.
Thus, by identifying useful inter-rule correlations, \ac{VNC} is able to model complex patterns in the knowledge graph and facilitate embedding learning.
\begin{table}
  \caption{
Ablation studies on FB15K Subject-10.}
  \label{Tab:Ablation studies on FB15K Subject-10}
    \begin{tabular}{l @{~} ccccc}  
    \toprule
    Model &MR &MRR &Hits@10 &Hits@3& Hits@1\\
    \midrule
    VNC & \textbf{151} & \textbf{54.3} & \textbf{75.9} & \textbf{60.8} & \textbf{41.6} \\
    \midrule
    \mbox{No rules} & 251 &40.9 &61.9 &  47.3 &31.5\\
    \mbox{Hard rules} & 192 & 45.2 & 67.6 & 52.6 & 35.4\\
    \mbox{Soft rules} & 164 & 53.3 & 74.2 & 58.5 & 40.1\\
    \midrule
    \mbox{VNC (AMIE+)} & 191 & 48.9 &71.3 &55.6 &37.2 \\
    \mbox{VNC (IterE)} & 172 & 52.5 &73.7 & 58.8 &39.7 \\
    \bottomrule
    \end{tabular}
\end{table}
\begin{table}[h]
  \caption{
Influence of the decoders on FB15K Subject-10.}
  \label{Tab:Influences of different decoders on FB15K Subject-10}
  \resizebox{\linewidth}{!}{
    \begin{tabular}{l ccccc}  
   \toprule
    Decoder &MR &MRR &Hits@10 &Hits@3 & Hits@1\\        
    \midrule
    VN network & 175 & 46.3 & 70.1 & 52.6 & 34.5\\
    \midrule
    VNC (TransE) & 204 &47.4 &71.2 &53.6 &34.5\\
    VNC (ConvE) &171 &53.1 &74.6 &59.8 &41.2\\
    VNC (Analogy) & 163 &52.9 &74.8 &60.1 &40.7\\
    VNC (DistMult) & \textbf{151} & \textbf{54.3} & \textbf{75.9} & \textbf{60.8} & \textbf{41.6}\\
    \bottomrule
    \end{tabular}
    }
\end{table}
\begin{table}[h]
\small
\caption{
Examples of inter-rule correlations on YAGO37. 
}
\label{Tab: case study}
 \resizebox{\linewidth}{!}{ 
    \begin{tabular}{l l}  
   \toprule
	\textbf{Soft label} & $s($\emph{George}, \emph{directed}, \emph{Cass\ Timberlane}$) = 0.98$\\
	\textbf{Logic rule} & $($\emph{George}, \emph{directed}, \emph{Young Bess}$)\rightarrow$\\
	&$ ($\emph{George}, \emph{created}, \emph{Young Bess}$)$
	\\ 
    \textbf{Incomplete rule}& $($\emph{George}, \emph{directed}, \emph{Cass Timberlane}$)\rightarrow$\\
    &$ ($\emph{George}, \emph{created}, \emph{Cass Timberlane}$)$
    \\
	\textbf{Inter-rule path}& $($\emph{Young Bess}, \emph{isLocatedIn}, \emph{United States}$)\land $\\
	& $($\emph{United States}, \emph{isLocatedIn}$^{-1}$, \emph{Cass Timberlane}$)$\\
	\midrule
	\textbf{Soft label} & $s($\emph{Sigmar}, \emph{isCitizenOf}, \emph{Germany}$) = 0.92$\\
	\textbf{Logic rule} & $($\emph{Thorbj{\o}rn}, \emph{hasChild}, \emph{Kjell}$)\land ($\emph{Kjell}, \emph{isCitizenOf}, \\
	& \emph{Norway}$)\rightarrow ($\emph{Thorbj{\o}rn}, \emph{isCitizenOf}, \emph{Norway}$)$\\ 
    \textbf{Incomplete rule}& $($\emph{Franz}, \emph{hasChild}, \emph{Sigmar}$)\land ($\emph{Sigmar}, \emph{isCitizenOf}, \\
    & \emph{Germany}$)\rightarrow ($\emph{Franz}, \emph{isCitizenOf}, \emph{Germany}$)$\\
	\textbf{Inter-rule path}& $($\emph{Germany}, \emph{hasNeighbor}, \emph{Denmark})\\ 
	& ${}\land ($\emph{Denmark}, \emph{dealWith}, \emph{Norway}$)$\\
    \bottomrule
    \end{tabular}
    }
\end{table}

\section{Conclusion and future Work}
\label{sec:Conclusion and future work}
In this paper, we focus on predicting missing facts for \acf{OOKG} entities.
Previous work for this task still suffers from two key challenges: identifying inter-rule correlations, and capturing the interactions within rule and embedding learning.
To address these problems, we propose a novel framework, named \ac{VNC}, that infers virtual neighbors for \ac{OOKG} entities by iteratively extracting logic rules and inter-rule correlations from knowledge graphs.
We conduct both link prediction and triple classification, and experimental results show that the proposed \ac{VNC} achieves state-of-the-art performance on four widely-used knowledge graphs.
Besides, the \ac{VNC} framework effectively alleviates the data sparsity problem, and is highly robust to the proportion of the unseen entities.

For future work, we plan to incorporate more kinds of complex patterns in knowledge graphs.
In addition, generalizing the \ac{VNC} framework to the unseen relations is also a promising direction.
\begin{acks}
This work was supported by the National Key R\&D Program of China (2020YFB1406704, 2022YFC3303004), the Natural Science Foundation of China (62272274, 61972234, 62072279, 62102234, 62202271), the Natural Science Foundation of Shandong Province (ZR2021QF129, ZR2022QF004), the Key Scientific and Technological Innovation Program of Shandong Province (2019JZZY010129), the Fundamental Research Funds of Shandong University, the China Scholarship Council under grant nr. 202206220085,
the Hybrid Intelligence Center, a 10-year program funded by the Dutch Ministry of Education, Culture and Science through the Netherlands Organization for Scientific Research, \url{https://hybrid-intelligence-centre.nl}, and project LESSEN with project number NWA.1389.20.183 of the research program NWA ORC 2020/21, which is (partly) financed by the Dutch Research Council (NWO).
\end{acks}

\clearpage
\bibliographystyle{ACM-Reference-Format}
\balance
\bibliography{references}

\end{document}